\documentclass[prd,twocolumn,superscriptaddress,nofootinbib,amsmath,amssymb]{revtex4-2}

\usepackage{graphicx}
\usepackage{dcolumn}
\usepackage{bm}
\usepackage{hyperref}
\usepackage{gensymb}
\usepackage{csquotes}
\usepackage{url}
\usepackage[dvipsnames]{xcolor}

\begin{document}

\preprint{APS/123-QED}

\title{Investigating starburst-driven neutrino emission from galaxies in the Great Observatories All-Sky LIRG Survey }

\author{Y. Merckx}
\email{yarno.merckx@vub.be}
\affiliation{Vrije Universiteit Brussel, Dienst ELEM, Pleinlaan 2, 1050 Brussels, Belgium}

\author{P. Correa}
\email{pabcorcam@gmail.com}
\affiliation{Vrije Universiteit Brussel, Dienst ELEM, Pleinlaan 2, 1050 Brussels, Belgium}\affiliation{Sorbonne Université, Université Paris Diderot, Sorbonne Paris Cité, CNRS, Laboratoire de Physique Nucleaire et de Hautes Energies (LPNHE), 4 place Jussieu, F-75252, Paris Cedex 5, France}

\author{K. D. de Vries}
\email{krijn.de.vries@vub.be}
\affiliation{Vrije Universiteit Brussel, Dienst ELEM, Pleinlaan 2, 1050 Brussels, Belgium}

 \author{K. Kotera}
 \email{kotera@iap.fr}
 \affiliation{Vrije Universiteit Brussel, Dienst ELEM, Pleinlaan 2, 1050 Brussels, Belgium}
 \affiliation{Sorbonne Université, CNRS, UMR 7095, Institut d’Astrophysique de Paris, 98 bis bd Arago, 75014 Paris, France}
 
\author{G. C. Privon}
\email{gprivon@nrao.edu}
\affiliation{National Radio Astronomy Observatory, 520 Edgemont Road, Charlottesville, Virginia 22903, USA}
\affiliation{Department of Astronomy, University of Florida, P.O. Box 112055, Gainesville, Florida 32611, USA}
\affiliation{Department of Astronomy, University of Virginia, 530 McCormick Road, Charlottesville, Virginia 22904, USA}

\author{N. van Eijndhoven}
\email{nick.van.eijndhoven@vub.be}
\affiliation{Vrije Universiteit Brussel, Dienst ELEM, Pleinlaan 2, 1050 Brussels, Belgium}

\date{\today}

\begin{abstract}

We present a phenomenological framework for starburst-driven neutrino production via proton-proton collisions and apply it to (ultra)luminous infrared galaxies (U/LIRGs) in the Great Observatories All-Sky LIRG Survey (GOALS). The framework relates the infrared luminosity of a GOALS galaxy, derived from consistently available \textit{Herschel Space Observatory} data, to the expected starburst-driven neutrino flux. The model parameters that define this relation can be estimated from multiwavelength data. We apply the framework in a case study to the LIRG NGC 3690 (Arp 299, Mrk 171) and compare the obtained neutrino fluxes to the current sensitivity of the IceCube Neutrino Observatory. Using our framework, we conclude that the neutrino emission in the LIRG NGC 1068, recently presented as the first steady IceCube neutrino point source, cannot be explained by a starburst-driven scenario and is therefore likely dominated by the active galactic nucleus in this galaxy. In addition to the single-source investigations, we also estimate the diffuse starburst-driven neutrino flux from GOALS galaxies and the total LIRG population over cosmic history.

\end{abstract}

\maketitle


\section{\label{sec:level1} Introduction}

In 1983, the \textit{Infrared Astronomical Satellite} (IRAS) was the first space-borne telescope to perform an all-sky survey at infrared (IR) wavelengths \cite{1984ApJ...278L...1N}. This novel view of the extragalactic sky revealed the existence of galaxies that emit most of their electromagnetic luminosity in the IR frequency range. Among these sources, so-called luminous infrared galaxies (LIRGs; $ 10^{11} L_{\odot} \leq L_{\mathrm{IR}} \equiv L_{\mathrm{IR}[8-1000\mu \mathrm{m}]} < 10^{12} L_{\odot}$) and ultraluminous infrared galaxies (ULIRGs; $L_{\mathrm{IR}} \geq $  10$^{12} L_{\odot}$) were discovered. Although these objects are relatively rare in the local Universe ($z<0.3)$\footnote{The number density of LIRGs, however, is larger than optically selected starburst galaxies and Seyfert galaxies at comparable redshifts and bolometric luminosities (e.g.$~$\cite{1987Soifer}).}, deep-sky IR observations show that the comoving number density of both LIRGs and ULIRGs has a positive redshift evolution. Furthermore, LIRGs appear to be more numerous than ULIRGs up to at least $z\sim$ 2 \cite{IRlum2005, Evolution, IRlum2010}.

Follow-up surveys revealed that LIRGs are systems of galaxies covering the entire evolutionary merger sequence, ranging from isolated galaxies, to early interacting systems, to advanced mergers \cite{HST1,Spitzer3}. This is opposed to ULIRGs which are nearly always involved in the final stages of a merger between two gas-rich galaxies \cite{Armus87,1996SSRv...77..303M,Sanders96}. The extreme IR output observed in U/LIRGs is a result of the dynamical nature of these objects. In the merger process, gas and dust are funneled toward the central $\sim$100 pc of the interacting galaxies, thereby triggering intense star formation ($\sim$10$-$100 M$_{\odot}$ yr$^{-1}$) \cite{SFR1,SFR2}. The strong radiation fields emerging from the newly formed stars heat thick layers of dust accumulated from active star formation, which reradiate the energy in the IR regime. This mechanism generally explains the elevated IR output of U/LIRGs. However, an additional contribution is expected from gas accretion onto a central supermassive black hole with growing evidence suggesting that they inhabit all massive galaxies (e.g.~\cite{SMBHintro}). This accretion can result in relativistic outflows of matter perpendicular to the plane of accretion. This elevated state of activity is known as an active galactic nucleus (AGN). Reprocessed high-energy emission from AGN activity can (significantly) contribute to the IR output of U/LIRGs (e.g.$~$\cite{AGNIR1,AGNIR2}). 
\\ 

The Great Observatories All-Sky LIRG Survey\footnote{Available at  \url{goals.ipac.caltech.edu/}.} (GOALS) aims to fully characterize the diversity of properties observed in a large, statistically significant sample of the nearest ($z<0.088$) U/LIRGs (Sec.~\ref{sec1}) \cite{GOALS}. This sample covers all galaxy-interaction stages \cite{Spitzer3,Morphology}. Moreover, GOALS galaxies span the full range of nuclear types, i.e.$~$type-1 and type-2 AGN, LINERs, and pure starbursts. 

GOALS combines data from space-borne facilities such as the \textit{Spitzer Space Telescope} \cite{2004ApJS..154....1W,Spitzer1,Spitzer2,Spitzer3,Spitzer4,Spitzer5} and \textit{Herschel Space Observatory} \cite{2010A&A...518L...1P,Herschel1,Herschel2,Herschel3, Herschel4, Herschel5,Herschel6,HerschelChu} at mid-IR and far-IR wavelengths, the \textit{Hubble Space Telescope} which observes near-IR and optical emission from the Universe \cite{HST1,HST2}, the \textit{Galaxy Evolution Explorer} (GALEX) UV telescope \cite{GALEX,GOALSUVFIR}, and the \textit{Chandra X-ray Observatory} operating in the x-ray frequency band \cite{C1,C2}. Recently, the \textit{James Webb Space Telescope} (JWST) was used for the first time to observe GOALS LIRGs with unprecedented resolution \cite{JWST1,JWST2,JWST3, JWST4, JWST5, JWST6,JWST7, JWST8}. In addition, GOALS objects are also targeted by ground-based observatories such as the radio and submillimeter telescopes \textit{Very Large Array} (VLA) \cite{VLA} and the \textit{Atacama Large Millimeter/submillimeter Array} (ALMA) \cite{ALMA} and large optical-IR facilities such as the Keck Telescopes \cite{2014ApJ...784...70M, Medling15,NuclearfeedbackLIRG}. The multiwavelength data from these space-borne and ground-based observatories are combined in comprehensive imaging and spectroscopic surveys. In this work, we take the first step towards expanding GOALS from a multiwavelength to a multi-messenger survey by investigating high-energy neutrino emission from these galaxies. 
\\

High-energy cosmic neutrinos were first discovered in 2013 with the 1-km$^3$ IceCube Neutrino Observatory buried deep within the ice at the South Pole \cite{IceCube2013}. To date, a diffuse astrophysical neutrino flux has been observed via various independent IceCube analyses \cite{Track1,Track2,Track3,Track4,Track5}. However, the sources of these cosmic neutrinos remain largely unknown. The IceCube Collaboration has performed several searches in order to identify the origin of the astrophysical neutrino flux \cite{ICRCicecube}. Such analyses typically target astrophysical muon neutrinos ($\nu_\mu$) and antineutrinos ($\bar{\nu}_{\mu}$).\footnote{IceCube cannot distinguish neutrinos from antineutrinos, except in the specific case of the Glashow resonance \cite{GlashowReso}. Therefore, the term \enquote{neutrinos} is used in this work to refer to both neutrinos and antineutrinos.}~Upon collision with ice nuclei, muon neutrinos can produce muons via charged-current interactions. These muons leave track-like Cherenkov signatures in the detector, allowing to reconstruct the incoming direction of the neutrino with an angular resolution $\lesssim$ 1$\degree$ for a muon energy $\gtrsim$ 1 TeV \cite{2020PhRvL.124e1103A}. One of the more generic analyses aims to identify steady point sources in a time-integrated sky scan by looking for spatial clustering of neutrino events on the sky. Recently, such a scan of the Northern Hemisphere, combined with a dedicated source-catalog search, revealed significant evidence (4.2$\sigma$) for neutrinos originating from the direction of the GOALS LIRG NGC 1068 \cite{Evidence1068}, which contains an enshrouded AGN surrounded by a starburst ring. Other evidence for a specific source was reported by IceCube after the spatial and temporal correlation between an IceCube neutrino event and the gamma-ray flaring blazar TXS 0506+056 \cite{TXS,ArchivalTXS}. Both sources contribute no more than about 1$\%$ to the diffuse neutrino flux in the energy ranges within which they were observed. As such, the origin of the diffuse flux remains largely unidentified. Nevertheless, diffuse multi-messenger observations of both neutrinos and gamma rays hint toward gamma-ray opaque neutrino sources (e.g.$~$\cite{Hidden1, Hidden2, ppobscured}). 
\\

Galaxies in the GOALS sample are characterized by a large amount of enshrouding matter and an enormous energy budget, driven by vigorous star formation and AGN activity. These two key features, in combination with the proximity of the sources, make U/LIRGs excellent candidate neutrino sources. A recent IceCube study sought neutrinos from the population of ULIRGs in particular, although null results were reported \cite{Pcor}. This allowed the authors to set upper limits on the contribution of the entire ULIRG population to the diffuse neutrino flux observed by IceCube. However, LIRGs show similar star-forming properties as ULIRGs and are $\sim$10-50 times more numerous than ULIRGs at any given redshift (e.g.~\cite{IRlum2005}). Therefore, it is crucial to investigate the contribution of \textit{both} LIRGs and ULIRGs to the IceCube neutrino flux. 

In this work, we present a phenomenological framework for starburst-driven neutrino production in starburst galaxies and apply this framework to the GOALS sample. In Sec.~\ref{sec1}, the GOALS sample is introduced, and Sec.~\ref{sec2} motivates these galaxies as candidate high-energy neutrino sources. Then, we construct a starburst-driven neutrino production framework in Sec.~\ref{framework}. Subsequently, this framework is applied to the LIRG NGC 3690 (also known as Arp 299 and Mrk 171) in Sec.~\ref{casestudy}. Finally, in Sec.~\ref{diffuseflux}, we use the framework to estimate the diffuse neutrino flux expected from the GOALS sample and the total LIRG population over cosmic history.

\section{The GOALS sample}\label{sec1}

\begin{figure}
\includegraphics[width=0.45\textwidth]{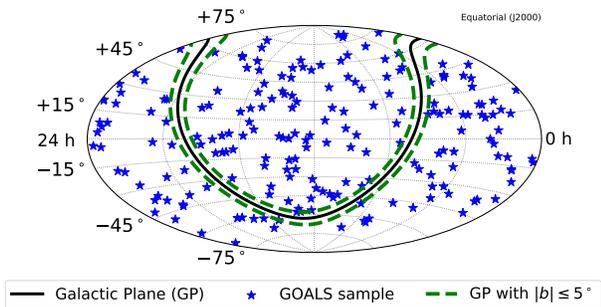}
\caption{\label{fig:skymap} Sky distribution of the GOALS sample. The solid line indicates the Galactic plane and the dashed lines indicate a band with $|b| \leq 5^{\degree}$.}
\end{figure}

The GOALS sample consists of 180 LIRGs and 22 ULIRGs with a median redshift of  $\langle z \rangle$ = 0.0212 \cite{GOALS}. The closest source in the sample is located at $z_{\mathrm{min}}$ = 0.0030 and the most distant one at $z_{\mathrm{max}}$ = 0.0876. The distribution of the GOALS sample on the sky is shown in Fig.~\ref{fig:skymap}. GOALS objects were originally selected from the IRAS Revised Bright Galaxy Sample (RBGS \cite{Sanders2003}) as sources with a luminosity threshold of $L_{\mathrm{IR,IRAS}} \geq 10^{11}L_{\odot}$. The RBGS consists of a complete flux-limited sample of 629 galaxies that have an IRAS 60-$\mu$m flux density $S_{60 \mu \mathrm{m}, \mathrm{IRAS}} > $ 5.24 Jy and Galactic latitude $|b| > 5^{\degree}$ \cite{Sanders2003}. This cut on the Galactic latitude is shown by the dashed lines in Fig.~\ref{fig:skymap}. 

GOALS is fundamentally based on observations of IRAS, which had a relatively low angular resolution between $\sim$0.5' at 12 $\mu$m and $\sim$2' at 100 $\mu$m \cite{IRASreso}. Therefore, the IRAS emission for a single GOALS object may correspond to the cumulative emission of individual galaxies in an interacting system. However, the framework presented in this work (Sec.~\ref{framework}) models neutrino production in the cores of U/LIRGs based on electromagnetic emission from those regions. Therefore, the IR luminosity of each galaxy in the interacting system is of interest, rather than the total IRAS IR luminosity of the system\footnote{Note that neutrino telescopes such as IceCube, with an angular resolution of the order of 1$\degree$, cannot resolve individual galaxies in interacting systems.}. In what follows, it is described how these individual IR luminosities were obtained by GOALS, as these will be used to trace starburst-driven neutrino production. We also discuss the contribution of AGN to the IR luminosities of the targeted galaxies.

\subsection{Individual IR luminosity \label{DisentangledIR}}

The \textit{Spitzer Space Telescope} (2003), one of IRAS' successors with a higher angular resolution, allowed to spatially disentangle galaxies within the same U/LIRG system. This yields more than 290 individual galaxies for the GOALS sample. Only a fraction of these galaxies were targeted by the \textit{Photodetecting Array Camera and Spectrometer} (PACS) onboard the \textit{Herschel Space Observatory} (2009). For U/LIRGs consisting of two or more galaxies, \textit{Herschel} only targeted the dimmer companion galaxies if their contribution to the total 24-$\mu$m flux-density ratio in the \textit{Multiband Imaging Photometer for Spitzer} (MIPS) exceeded 1:5 with respect to the brightest galaxy in the system. For those U/LIRG constituents that were targeted by \textit{Herschel}, the individual IR luminosity of a galaxy ($L_{\mathrm{IR, individual}}$) is obtained by applying a scaling factor to the IRAS luminosity of the system in which that galaxy resides ($L_{\mathrm{IR,IRAS}}$). The scaling factor is computed by taking the ratio between the continuum flux density detected for the individual galaxy, evaluated at 63 $\mu \mathrm{m}$ in the PACS spectrum ($S_{63 \mu \mathrm{m},\mathrm{PACS}}$), and the IRAS 60-$\mu \mathrm{m}$ flux density of the whole system ($S_{60\mu \mathrm{m} ,\mathrm{IRAS}}$). The IR luminosity of a disentangled component in a U/LIRG is then computed as

\begin{equation}
     L_{\mathrm{IR, individual}} = \frac{S_{63 \mu \mathrm{m},\mathrm{PACS}}}{S_{60\mu \mathrm{m},\mathrm{IRAS}}} \cdot L_{\mathrm{IR,IRAS}}~. 
\end{equation}

\begin{figure}
    \centering
    \includegraphics[width=0.48\textwidth]{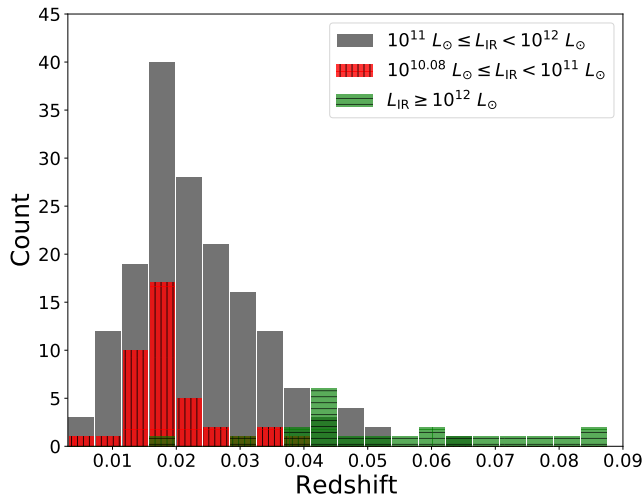}
    \caption{Redshift distributions of the 229 individual GOALS galaxies targeted in this work, consisting of 40 galaxies with $10^{10.08} L_{\odot} \leq L_{\mathrm{IR}} < 10^{11} L_{\odot}$, 167 galaxies with $ 10^{11} L_{\odot} \leq L_{\mathrm{IR}}< 10^{12} L_{\odot}$ (LIRGs), and 22 galaxies with $L_{\mathrm{IR}} \geq $  10$^{12} L_{\odot}$ (ULIRGs).}
    \label{splitredshiftdistr}
\end{figure}

\noindent This leads to individual IR luminosities for 229 GOALS galaxies, consisting of 40 galaxies with $10^{10.08} L_{\odot} \leq L_{\mathrm{IR}} < 10^{11} L_{\odot}$, 167 galaxies with $ 10^{11} L_{\odot} \leq L_{\mathrm{IR}}< 10^{12} L_{\odot}$, and 22 galaxies with $L_{\mathrm{IR}} \geq $  10$^{12} L_{\odot}$ \cite{HerschelSantos}. These individual IR luminosities will be used in Sec.~\ref{casestudy} and Sec.~\ref{diffuseflux} to trace the starburst-driven neutrino production in the respective sources. The redshift distributions of the three galaxy groups are shown in Fig.~\ref{splitredshiftdistr}. It is noted that the LIRGs are observed over the whole redshift range while the closest ULIRG Arp 220 is located at $z \sim 0.018$.

\subsection{AGN contribution to the IR luminosity \label{alphagn}} 

In this work, we focus on the starburst-driven neutrino emission in U/LIRGs which we trace via the observed IR luminosity. However, a significant fraction of the IR luminosity could be generated by AGN activity. This should be taken into account in order to not overestimate the starburst-driven neutrino flux. 

In \cite{HerschelSantos}, the average AGN contribution to the total bolometric luminosity, $\langle\alpha_{\mathrm{AGN}} \rangle \in [0,1]$, is presented for each of the galaxies introduced in Sec.~\ref{DisentangledIR}. By making use of low-resolution spectral measurements obtained with the \textit{InfraRed Spectrograph} (IRS) onboard \textit{Spitzer}, the $\langle\alpha_{\mathrm{AGN}} \rangle$-values were computed from a number of independent estimators such as the line ratios [Ne V]/[Ne II] and [O IV]/[Ne II], the mid-IR continuum slope, and the equivalent width of polycyclic aromatic hydrocarbon emission bands \cite{HerschelSantos}. Since for U/LIRGs the total IR luminosity approximates the total bolometric luminosity ($L_{\rm bol}$) \cite{SEDLIRG}, it follows that the IR luminosity can be corrected for AGN activity by multiplying it with the factor $1-\langle \alpha_{\rm AGN} \rangle$. This is used in Sec.~\ref{casestudy} and Sec.~\ref{diffuseflux} to properly estimate the starburst-driven neutrino flux expected from GOALS galaxies. 

The physical area of a galaxy probed by IRS observations depends on the angular scale covered by the short-low slit, $\sim4"\times4"$ \cite{Spitzer3}. Because of this limitation, the estimated $\langle \alpha_{\mathrm{AGN}}\rangle$-values are representative of an entire galaxy only for sources at luminosity distances $D_L \gtrsim$ 50$-$100 Mpc. Therefore, as noted by GOALS in \cite{HerschelSantos}, the galaxy-wide $\langle\alpha_{\mathrm{AGN}} \rangle$-values are to be interpreted as upper limits for more nearby sources. The most prominent example of this is NGC 1068 with a reported value $\langle\alpha_{\mathrm{AGN}} \rangle$ = 1. This galaxy is the closest Seyfert II galaxy to Earth, located at $D_L \sim 15.9$ Mpc. Only two other sources within 50 Mpc have  $\langle\alpha_{\mathrm{AGN}} \rangle >$  0.3, i.e.$~$the LIRGs NGC 1365 and NGC 4418. As the $\langle \alpha_{\mathrm{AGN}} \rangle$-value of these sources are only tracing the inner central part of the galaxy, it follows that, when considering the full system, these sources potentially have a smaller $\langle \alpha_{\mathrm{AGN}} \rangle$. 

\begin{figure}
    \centering
    \includegraphics[width=0.45\textwidth]{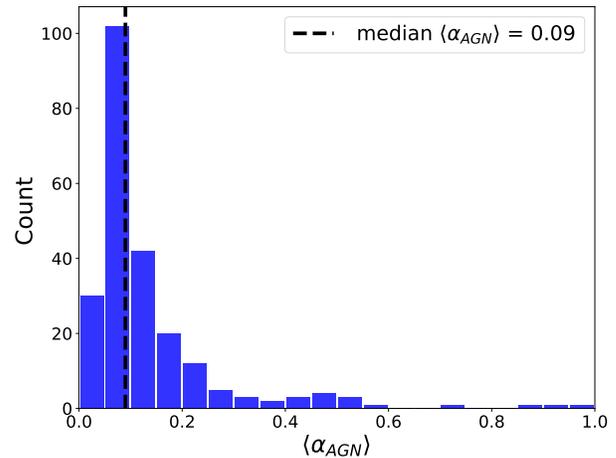}
    \caption{Distribution of the average AGN contribution to the bolometric luminosity for individual galaxies in the GOALS sample \cite{HerschelSantos}. The median of the sample is $\langle \alpha_{\mathrm{AGN}} \rangle$ = 0.09.}
    \label{agnboldistr}
\end{figure}

The distribution of $\langle\alpha_{\mathrm{AGN}} \rangle$-values in the GOALS sample is shown in Fig.~\ref{agnboldistr}. This distribution shows that in the majority of GOALS galaxies the AGN has a secondary contribution to the total bolometric luminosity. However, some of the sources show large $\langle\alpha_{\mathrm{AGN}} \rangle$-values, i.e.~14$\%$ have $\langle \alpha_{\mathrm{AGN}} \rangle > 0.2 $ and 3$\%$ have $\langle \alpha_{\mathrm{AGN}} \rangle> 0.5$ (AGN dominates over star formation). This implies, with a good consistency among different estimators, that only for 3$\%$ of the local U/LIRGs an AGN is the dominant power source.

\section{Motivating GOALS galaxies as candidate neutrino sources }\label{sec2}

The majority of GOALS objects are galaxies participating in a dynamical interaction. Such interactions allow for large amounts of dust and gas to be funneled from kpc-scales to the innermost regions of the merging galaxies. This generates pressure waves in the central region and thereby triggers intense star formation. Such starburst regions consist of short-lived, hot, massive stars that emit strong UV radiation fields. This radiation heats the enshrouding matter in which the stars were formed, and this heat is subsequently reradiated as thermal IR emission. Therefore, IR luminosity and starburst activity are intimately connected in dust-obscured environments, such as the nuclei of U/LIRGs. The massive stars in the starburst region burn significantly faster through the hydrogen phase than low-mass stars. This results in an increased rate of core-collapse supernova events for stars with masses $\gtrsim$ 8 $M_{\odot}$. The supernova rate can be $\sim$10$-$100 times larger than for normal star-forming galaxies such as the Milky Way. During a supernova explosion, the outer layer of the star is ejected with a kinetic energy of $\sim$10$^{51}$ erg (e.g.~\cite{Explosionenergy}). Upon collision, these supersonic ejecta drive strong shock waves with large Mach numbers in the surrounding medium. Particles can be accelerated along these shocks via diffusive shock acceleration \cite{DSA1,DSA2}, which is based on the first-order Fermi mechanism \cite{1954ApJ...119....1F}. A fraction of the accelerated hadrons are expected to reach the threshold energy to interact with the radiation fields and merger-enhanced matter in the starburst region via photohadronic and inelastic hadronuclear interactions, respectively. Both interactions produce, along with other particles, charged ($\pi^{\pm}$) and neutral ($\pi^0$) pions. The charged pions decay to high-energy neutrinos,  $\pi^{\pm} \to \mu~\nu_{\mu} \to e~\nu_e \nu_\mu \nu_\mu$ (e.g.~\cite{98Waxman,Peretti2}), and the neutral pions to gamma rays, $\pi^0 \to \gamma \gamma $ (e.g.~\cite{WBstarburst,StarburstFR}). Thus far, eleven star-forming galaxies, including four GOALS U/LIRGs, have been identified as gamma-ray sources \cite{gammaraySFG}. 

As already noted, AGN activity is found in several of the GOALS galaxies. The unified AGN model (see e.g.$~$\cite{UnifiedAGN} for a review) states that the AGN is powered by accretion of matter onto a supermassive black hole. In U/LIRGs, this is triggered and sustained via the flow of gas and dust toward the central region as the merger progresses. This forms an accretion disk which emits high-energy UV/optical radiation and can result in relativistic outflows of ionized matter perpendicular to this disk. Particle acceleration is possible both in the relativistic jets and in the thermal plasma above the accretion disk (see e.g.$~$\cite{AGNmurase}). The hadrons accelerated in this way can interact with the strong radiation fields in the AGN vicinity via photohadronic collisions \cite{AGNmurase}. Moreover, the accelerated hadrons can inelastically collide with thermal hadrons in, for example, the accretion disk, the dusty torus surrounding the accretion disk, or in a cloud in the line of sight of the out-flowing jet \cite{ppobscured}. Much as in the starburst-driven scenario, these collisions produce high-energy neutrinos. 
\\

A subset of the GOALS U/LIRGs host extremely compact and dusty nuclei in the central 100 pc, known as compact obscured nuclei (CONs, e.g.~\cite{CON1,CON2}). These CONs  can generate a significant fraction of the total IR output of the galaxy. The high column density ($N_\mathrm{H} \gtrsim 10^{25}$ cm$^{-2}$) in CONs implies a dust optical thickness above unity up to at least far-IR wavelengths. The most obscured systems only become optically thin at submillimeter and radio wavelengths, e.g.~the ULIRG Arp 220 \cite{Arp220CON}, and the LIRGs IC 860 \cite{IC860} and NGC 4418 \cite{4418CON}. Consequently, the source powering these CONs remains unknown. They could be driven by hidden AGN activity, a nuclear starburst with a top-heavy initial mass function, or a combination of both \cite{CON1}. If AGN activity is at the origin, then CONs could be the result of rapid accretion onto a supermassive black hole, surrounded by extreme column densities. In any case, if hadronic acceleration occurs, a large cosmic-ray density is expected in a compact and obscured region. Such an extreme environment provides favorable conditions for high-energy neutrino production, while also significantly attenuating gamma rays. The latter is of interest as diffuse observations of both neutrinos and gamma rays hint toward gamma-ray opaque neutrino sources (e.g.~\cite{Hidden1}). 
\\

As outlined in this section, obscured star formation and AGN activity, traced by strong IR emission, provide favorable conditions for high-energy neutrino production. In this work, the focus lies solely on high-energy neutrino production driven by supernova activity. This does not exclude other starburst-driven activity, e.g.~newborn pulsars \cite{Pulsars}, and AGN-related processes as interesting neutrino sources. The recent detection of high-energy neutrinos from the direction of the LIRG NGC 1068, for example, points toward the AGN being the dominant source of neutrinos in this galaxy (\cite{Evidence1068} and see Sec.~\ref{PerandDiffuseGOALS}). Additionally, we note that tidal disruption events (TDEs), i.e.~when a star is tidally disrupted by the gravitational pull of a supermassive black hole (see \cite{TDErev} for a review), are candidate sources of high-energy neutrinos \cite{TDEneutrinos}. U/LIRGs could have an increased rate of TDEs as a result of the amplified star-formation rate in their nuclear regions.

In terms of neutrino-production channels, our starburst-driven model does not take into account photohadronic interactions. For U/LIRGs, the target radiation field is likely dominated by IR emission. The threshold energy for a cosmic ray to interact with a near-IR background photon ($\sim$1 eV) is of the order of 100 PeV. The threshold energy for the cosmic ray is even larger for a target field dominated by far-IR radiation. Such extreme cosmic-ray energies are unlikely to be produced efficiently by the starburst activity considered in this work. Finally, as the fraction of elements heavier than protons is subdominant both in the acceleration region and the target interstellar medium (ISM), we opt to only consider high-energy neutrino production in proton-proton (pp) collisions.

\section{\label{sec:mathfr}Neutrino production framework \label{framework}}

Each (circum)nuclear starburst region in the GOALS sample has a related core-collapse supernova rate $(\mathcal{R}_{\mathrm{SN}}$, typically in units [yr$^{-1}$]). This rate drives the  high-energy proton injection rate $\left(Q_\mathrm{p}~[(\mathrm{GeV}/\mathrm{c})^{-3}~\mathrm{cm}^{-3}~\mathrm{s}^{-1}]\right)$ in the starburst volume $\left(V_{\mathrm{SBR}}~[\mathrm{pc}^{3}]\right)$. After injection, these cosmic rays reside for an average time $\left(\tau~[\mathrm{s}]\right)$ in the volume. The interplay between the proton injection rate and the residence time determines the distribution of high-energy proton momenta $\left(\mathcal{F}_\mathrm{p}~[(\mathrm{GeV}/\mathrm{c})^{-3}~\mathrm{cm}^{-3}]\right)$. The latter provides information on the available energy budget for charged pion production, which is required to compute the neutrino production rate $\left(q_{\nu}~[\mathrm{GeV}^{-1}~\mathrm{cm}^{-3}~\mathrm{s}^{-1}]\right)$. Finally, the expected neutrino flux $\left(\Phi_{\nu}~[\mathrm{GeV}^{-1}~\mathrm{cm}^{-2}~\mathrm{s}^{-1}]\right)$ is found by integrating the neutrino production rate over the starburst volume and taking into account the luminosity  distance to the source $\left(D_{L}~[\mathrm{Mpc}]\right)$.

In the following, we construct a phenomenological framework based on the above-mentioned parameters to compute a per-source starburst-driven neutrino flux for all GOALS galaxies. Our framework builds on the model of cosmic-ray transport in starburst nuclei presented in \cite{Peretti1} and previous investigations of starburst regions as potential sources of high-energy neutrinos (e.g.~\cite{nusb1,nusb2,nusb3,sbps,SBGstrikeback}). We contribute to these models by placing them in the context of local U/LIRGs as candidate neutrino sources. Moreover, our framework provides an approach to estimate the cosmic-ray luminosity per source via the IR luminosity, the AGN contribution to the IR luminosity, and the initial mass function of the studied region.

\subsection{\label{subsec:supernova rate}Supernova rate in the starburst region}

Optical emission from a supernova explosion is known to outshine entire galaxies. Given the large amounts of obscuring matter in GOALS galaxies, one cannot rely on optical counting to compute supernova rates in these galaxies. Nevertheless, there are numerous electromagnetic tracers (nearly) unaffected by obscuring matter that can be related to star-forming activity. There have been individual supernova counting experiments in U/LIRGs using near-IR emission (e.g.~\cite{SNIC883}) and radio emission (e.g.~\cite{SNArp220}). It is, however, not feasible to do this for the entire GOALS sample. As such, we opt to relate the total IR luminosity of a galaxy to the star-formation rate and subsequently relate this star-formation rate to the core-collapse supernova rate via scaling relations. We also take into account that part of the total IR luminosity could be generated by AGN activity and regions outside the central $\sim$100 pc of interest. This allows us to estimate the AGN-corrected nuclear supernova rate. 

We use the IR emission as it is available for all GOALS galaxies (Sec.~\ref{sec1}), and as such allows us to estimate the diffuse neutrino flux for the GOALS sample in Sec.~\ref{diffuseflux}. Radio emission is also an interesting tracer in this context as it is a direct tracer of particle acceleration. However, such data is not uniformly available for all GOALS U/LIRGs. Moreover, it is not straightforward to connect the relativistic electron population, traced by synchrotron emission, with any associated proton population.

\subsubsection{Calibrating the supernova rate to the IR luminosity} 
The bolometric luminosity of young stellar populations is dominated by massive, short-lived, UV-bright stars. Therefore, the UV luminosity is a sensitive probe for recent star formation. The presence of obscuring matter can lead to severe attenuation of UV photons, which are reprocessed into thermal emission. IR and UV emission can therefore be used to trace obscured and unobscured star formation, respectively. A study of 135 GOALS U/LIRGs shows that the far-UV measured by GALEX contributes an average of $\sim$4 $\%$ to the overall star-formation rate \cite{GOALSUVFIR}. Therefore, we opt to only use the IR luminosity to trace the star-formation rate in GOALS U/LIRGs. 

In general, each tracer of the star-formation rate is mapped back to star formation via two main relations, i.e.$~$the initial mass function (IMF) and the star-formation history (SFH). The IMF describes the mass distribution of a population of stars at formation time within a volume of space. The IMF is typically well-described by a power law of the form $\zeta(m) \propto m^{- \beta}$, with $m$ the stellar mass and $\beta(m)$ the power-law index (see e.g.$~$\cite{IMFreview} for a review). The latter can have different values for different stellar-mass ranges. The SFH describes how the star-formation rate evolved over time. Given an IMF, SFH, and a stellar-evolution model, it is possible to determine from simulations the calibration factor ($A_{\mathrm{IR}}$) that relates star-formation rate to IR luminosity. 

To obtain the calibration factor $A_{\mathrm{IR}}$ for different IMFs, we use the web-based software \textit{Starburst99}\footnote{Available at \url{www.stsci.edu/science/starburst99/docs/default.htm}.} (SB99)  \cite{SB9999,SB9905, SB9910,SB9914}. This software allows to model spectrophotometric properties of star-forming galaxies, such as the time-dependent spectral energy distribution (SED) of a stellar population. Following the procedure outlined in \cite{Murphy2011} (M11 from here on), we assume that the entire Balmer continuum, i.e.$~$stellar UV emission between 912 \AA~$<$ $\lambda$ $<$ 3646 \AA, is absorbed by dust and reradiated as optically-thin thermal IR emission. This implies that the IR luminosity due to reprocessed stellar emission, $L_{\mathrm{IR,SED}}$, is obtained by integrating the Balmer range of the simulated SED. As such, the calibration factor is defined as
\begin{equation}
      \Big(\frac{\mathrm{SFR}_\mathrm{SB99}}{M_{\odot}~\mathrm{yr}^{-1}}\Big) = A_{\mathrm{IR}} \cdot \Big(\frac{L_{\mathrm{IR,SED}}}{\mathrm{erg~s^{-1}}}\Big)~,
\end{equation}

\noindent with $\mathrm{SFR}_\mathrm{SB99}$ the star-formation rate used as input to run the SB99 simulation. The value of the calibration factor in M11, assuming a Kroupa IMF (see Sec.~\ref{computingcalib}), solar metallicity, and a constant SFH, is $A_{\mathrm{IR}} = 3.88 \times 10^{-44}$. In a follow-up study (\cite{Murphy2012}, M12 from here on), an empirical approach resulted in a linear relation between the star-formation rate and $L_{\mathrm{IR}}$ resulting in $A_{\mathrm{IR}} = 3.15 \times 10^{-44}$. This empirical relation is quoted to be reliable within a factor of two. The calibration factor obtained in M11 via SB99 is therefore consistent with the empirical calibration factor in M12. In this work, $A_{\mathrm{IR}}$ is computed in Sec.~\ref{computingcalib} for various types of IMFs using SB99, including the IMF of M11 for comparison.

SB99 also provides the total supernova rate as a function of time for the same stellar population. This allows to compute a calibration between the supernova rate and star-formation rate as
\begin{equation}
    \Big(\frac{\mathcal{R}_{\mathrm{SN,SB99}}}{\mathrm{yr}^{-1}}\Big) = A_{\mathrm{SFR}} \cdot  \Big(\frac{\mathrm{SFR}_\mathrm{SB99}}{M_{\odot}~\mathrm{yr}^{-1}}\Big)~,
\end{equation}

\noindent with $\mathcal{R}_{\mathrm{SN,SB99}}$ provided by SB99. Both calibration factors, $A_{\mathrm{IR}}$ and $A_{\mathrm{SFR}}$, are computed in a regime where the SED and supernova rate reach an equilibrium.

Combining both calibration factors gives $\Lambda_{\mathrm{IR}} = A_{\mathrm{SFR}}\cdot A_{\mathrm{IR}}$, such that the total supernova rate in the whole galaxy is estimated as 
\begin{equation}
    \Big(\frac{\mathcal{R}_{\mathrm{SN}}}{\mathrm{yr}^{-1}}\Big) = \Lambda_{\mathrm{IR}} \cdot  \Big(\frac{L_{\mathrm{IR}}}{\mathrm{erg~s^{-1}}}\Big)~,
    \label{totalsupernova}
\end{equation}

\noindent with $L_{\mathrm{IR}}$ the IR luminosity of that galaxy.

\subsubsection{\label{computingcalib}Computing the calibration factors}

To compute the value of the calibration factors $A_{\mathrm{IR}}$ and $A_{\mathrm{SFR}}$, the SB99 input parameters must be fixed. We consider a constant SFH and solar metallicity. As the parametrization of the IMF and its universality remains uncertain, it is not straightforward to select an appropriate IMF. We therefore investigate the effect of two different classes of IMFs. The first class consists of the Salpeter IMF (1953) \cite{1955ApJ...121..161S}, which has a single power-law exponent $\beta$ = 2.35, and a Kroupa IMF (2001) \cite{2001MNRAS.322..231K}, with $\beta_{\mathrm{low}} = 1.3$ for 0.1 $<$ $m/M_{\odot}$ $<$ 0.5 and $\beta_{\mathrm{high}} = 2.3$ for 0.5 $<$ $m/M_{\odot}$ $<$ 100. These so-called canonical IMFs are based on resolved stellar populations in the Milky Way and nearby galaxies.  For the second class, we consider two top-heavy IMFs. Such IMFs predict relatively more heavy-mass stars than expected from canonical IMFs. The interest in a top-heavy IMF for starburst regions is justified from both a theoretical and data-driven point of view. Theoretically, this is argued by the increased temperature in star-forming clouds due to the enhanced cosmic-ray density in starburst regions. This increase in temperature leads to a larger Jeans mass in the star-forming clouds, which suppresses the formation of low-mass stars  \cite{JeansMass2011}. This implies a change of the IMF shape toward a top-heavy IMF. In addition, high-resolution ALMA observations of nearby U/LIRGs suggest unusually low $^{13}$C/$^{18}$O isotope abundance ratios \cite{COtopheavy,COULIRG2019, IMFNature}. Short-lived massive stars ($\gtrsim$ 8 $M_{\odot}$) are the predominant source of $^{18}$O in the ISM while the $^{13}$C atom is convected into envelopes of long-lived, low-mass stars ($\lesssim$ 8 $M_{\odot}$). Therefore, unusually small values of the abundance ratio hints toward relatively more short-lived massive stars than expected. 

The SB99 simulations show that $\sim$60 Myr after the onset of star formation, the supernova rate stabilizes, assuming a constant SFH. Therefore, all calibrations were computed beyond this timestamp.  Table \ref{tab:imf} shows the computed calibration factors at 100 Myr for a Salpeter, Kroupa, and two top-heavy IMFs. The latter have the same low-mass exponent $\beta_{\mathrm{low}}$ = 1.3 as the canonical Kroupa IMF discussed earlier, but the high-mass exponent is taken to be $\beta_{\mathrm{high}} = 1.0$ and $\beta_{\mathrm{high}}$ = 1.5. An exponent such as the latter has been suggested to explain the reionization of the intergalactic medium at  $z \lesssim$ 11 \cite{THatIon}. The value $\beta_{\mathrm{high}} = 1.0$ is chosen as an arbitrary extreme case. First, it is concluded that $A_{\mathrm{IR}}$ obtained from the Kroupa IMF is a factor 1.27 larger than the value obtained in M11, which uses the same IMF. This increase is still consistent with the empirical calibration between star-formation rate and total IR luminosity presented in M12. Second, comparing the $A_{\mathrm{IR}}$ values obtained from all the investigated IMFs, it is concluded that $A_{\mathrm{IR}}$ is significantly lower for top-heavy IMFs as compared to the canonical IMFs. For a fixed IR luminosity, this results in a lower star-formation rate for top-heavy IMFs as opposed to the star-formation rate obtained from the canonical IMFs. This difference is, however, less prominent for the $\Lambda_{\mathrm{IR}}$ calibration factor. Compared to the canonical Kroupa IMF, the supernova rate for a fixed $L_{\mathrm{IR}} = 10^{11} L_{\odot}$ is a factor 1.31 lower for the top-heavy IMF with $\beta_{\mathrm{high}} = 1.5$ and a factor 1.75 lower for a top-heavy IMF with $\beta_{\mathrm{high}} = 1.0$. Although the total predicted supernova rate decreases for a top-heavy IMF, the average progenitor mass per supernova event is larger, and as a consequence also the average explosion energy per supernova event $E_{\mathrm{SN}}$ (see e.g.~\cite{2021Natur.589...29B}). This increase affects the supernova luminosity, i.e.$~\mathcal{L_{\mathrm{SN}}}$ =  $E_{\mathrm{SN}} \cdot \mathcal{R}_{\mathrm{SN}}$, which is required to compute the high-energy particle budget available for neutrino production. In the following section, it is discussed how this effect is taken into account in this work. 

\begin{table}
\caption{\label{tab:imf}
Calibration factors for different IMFs at 100 Myr, using a solar metallicity and constant SFH. The supernova rate $\mathcal{R}_{\mathrm{SN}}$ is computed for a fixed IR luminosity of $L_{\mathrm{IR}}$ = 10$^{11}$ $L_{\odot}$ via Eq.$~(\ref{totalsupernova})$. }
\begin{ruledtabular}
\begin{tabular}{ccccc}
 & $A_{\mathrm{IR}} \times 10^{44}$ & $A_{\mathrm{SFR}}$ & $\Lambda_{\mathrm{IR}}\times 10^{46}$ & $\mathcal{R}_{\mathrm{SN}}$ [yr$^{-1}$]\\
\hline
Salpeter & $7.48$ & 0.008 & $5.98$ & 0.23 \\
Kroupa & $4.93$ & 0.012 & $5.97$ & 0.23 \\
TH  $\beta_{\mathrm{high}} = 1.5$ & $1.54$	&  0.027  & 4.15 & 0.16   \\
TH $\beta_{\mathrm{high}} = 1.0$ & $1.24$ & 0.026 & 3.24 & 0.12	\\
\end{tabular}
\end{ruledtabular}
\end{table}

\subsubsection{The effect of the average supernova progenitor mass on the supernova luminosity \label{M}}

To account for the increase in average progenitor mass when considering top-heavy IMFs, we take the normal Kroupa IMF as a benchmark and fix the average energy per supernova event to $E_{\mathrm{SN, bm}}$ = 10$^{51}$ erg. We assume in this work that the explosion energy scales linearly with the average progenitor mass. Then, the average energy per supernova event for a different IMF is found by the scaling relation
\begin{equation}
    E_{\mathrm{SN}} = \frac{\langle M_{\mathrm{SN,IMF}} \rangle}{\langle M_{\mathrm{SN,bm}} \rangle} \cdot E_{\mathrm{SN,bm}} = \mathcal{M}\cdot E_{\mathrm{SN,bm}}~.
    \label{progenitormassformula}
\end{equation}

\noindent $\mathcal{M}$ is the fraction of the typical mass per supernova event for the chosen IMF, $\langle M_{\mathrm{SN,IMF}} \rangle$, over the typical mass per supernova event for the benchmark case, $\langle M_{\mathrm{SN,bm}} \rangle$. Based on our SB99 simulations, we find $\mathcal{M}$ = 1.37 for a top-heavy IMF with $\beta_{\mathrm{high}} = 1.5$ and $\mathcal{M}$ = 2.01 for the top-heavy IMF with $\beta_{\mathrm{high}} = 1.0$. Taking this factor into account, it is found that the supernova luminosity $\mathcal{L}_{\mathrm{SN}}$ for the top-heavy IMF with $\beta_{\mathrm{high}}$ = 1.5 and the benchmark case differ by $5 \%$. For the top-heavy IMF with $\beta_{\mathrm{high}}$ = 1.0 this correction gives a supernova luminosity which is $\sim$15 $\%$ larger than found for the canonical Kroupa IMF. 

\subsubsection{Correcting the IR luminosity for AGN contamination and extended IR emission \label{IRcorrections}}

The SB99 simulations do not take into account AGN activity. However, strong AGN activity in U/LIRGs could heat the matter in the (circum)nuclear region around the supermassive black hole. This heating can significantly contribute to the observed IR luminosity of its host galaxy. Therefore, using the total IR luminosity as tracer for the star-formation rate in the presence of a strong AGN can significantly overestimate the actual supernova rate. To correct the IR luminosity for this, the relative AGN contribution to the bolometric luminosity $\langle \alpha_{\mathrm{AGN}} \rangle$ (see Sec.~\ref{sec1}) is used. Doing this is justified, as per definition for U/LIRGs, $L_{\mathrm{bol}} \sim L_{\mathrm{IR}}$ (e.g.$~$\cite{SEDLIRG}). Furthermore, in this work, we are interested in the nuclear supernova rate in the central $\sim$100 pc. Therefore, the AGN-corrected IR luminosity of the nuclear region ($L_{\mathrm{IR, nuclear}}$) is required rather than the total IR luminosity of the galaxy. To take this into account, the factor $\mathcal{G}$ $\in$ [0,1] is introduced, which describes the amount of IR luminosity generated by nuclear starburst activity. As such, $L_{\mathrm{IR, nuclear}}$ = $\mathcal{G}\cdot \left(\left[1-\langle\alpha_{\mathrm{AGN}}\rangle\right]\cdot L_{\mathrm{IR}}\right)$. IR observations of U/LIRGs show that systems with larger $L_{\mathrm{IR}}$ tend to have a more centrally concentrated emission (e.g.$~$\cite{Spitzer1, Spitzer2}).  Therefore, $\mathcal{G}$ is likely to be closer to unity for systems with larger $L_{\mathrm{IR}}$. Targeted observations of four GOALS LIRGs show that $ \mathcal{G} \gtrsim 0.5$ for these galaxies \cite{SongLIRGs}. 

The nuclear AGN-corrected supernova rate per resolved galaxy in the GOALS sample is therefore calculated as

\begin{equation}
    \Big(\frac{\mathcal{R}_{\mathrm{SN}}}{\mathrm{yr}^{-1}}\Big) =  \Lambda_{\mathrm{IR}} \cdot \Bigg(\frac{\mathcal{G} \cdot \left[1-\langle\alpha_{\mathrm{AGN}}\rangle\right]\cdot L_{\mathrm{IR}}}{\mathrm{erg~s}^{-1}}\Bigg) ~.
    \label{RSNformula}
\end{equation}

Using Eq.~(\ref{RSNformula}), we can estimate the supernova rates in each of the 229 individual GOALS galaxies targeted in this work (see Sec.~\ref{sec1}). To do so, we use the IR luminosity of the galaxy and its corresponding $\langle \alpha_{\mathrm{AGN}}\rangle$-value, discussed in Sec.~\ref{sec1}. Then, for $\Lambda_{\mathrm{IR}} = 5.97 \times 10^{-46}$ (Table \ref{tab:imf}) and $\mathcal{G} = 1$ in all galaxies,  we find a median supernova rate of $\mathcal{R}_{\mathrm{SN}}$ = 0.43 yr$^{-1}$, a minimum supernova rate of $\mathcal{R}_{\mathrm{SN}}$ = 0.02 yr$^{-1}$, and a maximum supernova rate of $\mathcal{R}_{\mathrm{SN}}$ = 7.53 yr$^{-1}$.

\subsection{Proton injection rate}

Cosmic-ray acceleration via diffusive shock acceleration is expected along the forward shock in core-collapse supernova (CCSN) remnants. This mechanism gives rise to a power-law differential momentum distribution of accelerated particles. Therefore, a $p^{-\gamma_{\mathrm{SN}}}$ power-law relation between the proton injection rate ($Q_\mathrm{p}$) and the injected momentum $p$ is adopted. In addition, an exponential cutoff is considered at the maximum momentum $p_{\mathrm{max}}$ achieved in the acceleration process. The values of both $\gamma_{\mathrm{SN}}$ and $p_{\mathrm{max}}$ are discussed below. The total injection rate of high-energy protons per unit volume due to all CCSN in a (circum)nuclear starburst region is then expressed as

\begin{align}
Q_{\mathrm{p}} = \frac{N_{C}
}{V_{\mathrm{SBR}}} \left[\frac{p}{m_{\mathrm{p}}c}\right]^{-\gamma_{\mathrm{SN}}}e^{\frac{-p}{p_{\mathrm{max}}}}~.
\label{Qp}
\end{align}

    \noindent $N_{C}$ is the normalization constant to be fixed by the supernova rate in the starburst region, $V_{\mathrm{SBR}}$ is the volume of the region under consideration, and $m_\mathrm{p}$ is the proton mass. In the following sections, each of the parameters of Eq.$~(\ref{Qp})$ are discussed in more detail.

\subsubsection{Geometry of the starburst region\label{geometry}}

Hydrodynamic simulations of mergers between gas-rich galaxies predict the formation of nuclear gas disks on scales of $\sim$10$-$100 pc (e.g.$~$\cite{2008MmSAI..79.1284M}). Observational evidence for such gas disks in the nuclear regions of GOALS U/LIRGs is provided by a survey targeting 17 nearby U/LIRGs \cite{2014ApJ...784...70M}. Within this gas-disk configuration, stars are formed, which eventually explode as supernovae and thereby inject cosmic rays into the nuclear ISM. Based on these simulations and observations, we opt for a disk geometry to model the volume in which cosmic rays propagate. This disk is parametrized by a radius $R_{\mathrm{SBR}}$ and a scale height $H_{\mathrm{SBR}}$. This implies that the volume of the starburst region is computed as $V_{\mathrm{SBR}} = 2 H_{\mathrm{SBR}} \pi R^2_{\mathrm{SBR}}$, with $2H_{\mathrm{SBR}}$ the total thickness of the nuclear disk.

\subsubsection{Normalizing the injection rate to the cosmic-ray luminosity } 

The normalisation constant $N_C$ in Eq.$~$(\ref{Qp}) of the proton injection rate is determined by imposing
\begin{equation}
    \mathcal{L}_{\mathrm{CR}} = \int_{p_{\mathrm{min}}}^{p_{\mathrm{max}}} 4\pi p^2 \cdot N_{C} \cdot \left(\frac{p}{m_\mathrm{p} c}\right)^{-\gamma_{\mathrm{SN}}} \cdot ~ e^{\frac{-p}{p_{\mathrm{max}}}} \cdot  \mathcal{T}(p) ~\mathrm{d}p~.
\end{equation}

\noindent $\mathcal{T}(p) = \sqrt{p^2c^2+m_{\mathrm{p}}^2c^4}-m_{\mathrm{p}}c^2$ is the kinetic energy of a single cosmic-ray particle, and $\mathcal{L}_{\mathrm{CR}}$ is the total cosmic-ray luminosity due to CCSN activity in the nuclear starburst region. The minimum proton momentum $p_{\mathrm{min}}$ is fixed to\footnote{The value of $N_C$ is weakly dependent on the choice of $p_{\mathrm{min}}$ for $p_{\mathrm{min}}$ $\lesssim$ 0.1 GeV/$c$. } $p_{\mathrm{min}} = 0.1$ GeV/$c$ and the cosmic-ray luminosity $\mathcal{L}_{\mathrm{CR}}$ is computed as

\begin{equation}
\mathcal{L}_{\mathrm{CR}} = \eta_{\mathrm{SN}} \cdot \mathcal{R}_{\mathrm{SN}}\cdot  E_{\mathrm{SN}} = \eta_{\mathrm{tot}}\cdot L_{\mathrm{IR}}~.
\label{CRlumformula}
\end{equation}

\noindent The total CCSN rate $\mathcal{R}_{\mathrm{SN}}$ and kinetic energy output per supernova $E_{\mathrm{SN}}$ are computed as discussed in Sec.~\ref{subsec:supernova rate}. The conversion factor $\eta_{\mathrm{SN}}$ determines the amount of kinetic energy from the outflow that goes into the acceleration of cosmic-ray particles. The observed cosmic-ray spectrum at Earth up to $\sim$3 PeV can be explained with $\eta_{\mathrm{SN}} \simeq$ 0.10$-$0.30 for the bulk of the supernovae in the Milky Way (e.g.$~$\cite{Bykovreview}, \cite{9lives} for reviews). Moreover, kinetic simulations show that this energy transfer can be as much as $\eta_{\mathrm{SN}} \simeq $ 0.10$-$0.20  \cite{crefficiency}. This indicates that the conversion can be anything between $\eta_{\mathrm{SN}} \simeq$ 0.10$-$0.30 as long as predictions relying on this conversion factor are compatible with observations. The factor $\eta_{\mathrm{tot}}$ describes the fraction of IR luminosity which is related to the cosmic-ray luminosity due to starburst activity.

\subsubsection{\label{spectralindex} Spectral index of proton injection}

Diffusive shock acceleration in the presence of strong shock waves, such as those driven by supernova ejecta, predicts $Q_\mathrm{p} \propto p^{-4}$. However, observations of Galactic supernova events typically require softer spectra to model their gamma-ray spectra (e.g.$~$\cite{GammaSN,SNpevatron2}). The value of $\gamma_{\mathrm{SN}}$ for a GOALS galaxy can be estimated from the spectral index ($\Gamma$) of the starburst-driven hadronic gamma-ray spectrum of that galaxy (see e.g.~\cite{Peretti1,SBGstrikeback}). The value of $\Gamma$ can in turn be obtained by fitting the observed gamma-ray flux $\Phi_{\gamma}$ of a galaxy with a function of the form $\Phi_{\gamma} \propto E^{-\Gamma}$. Both spectral indices are related as $\Gamma = \gamma_{\mathrm{SN}} - 2$ because $\gamma_{\mathrm{SN}}$ is associated with momentum space and $\Gamma$ with energy space. As both gamma rays and neutrinos are expected to follow the same spectral shape, the spectral index $\gamma$ of the neutrino flux $\Phi_\nu$ can also be estimated from $\Gamma$. Note that $\Phi_\nu \propto E^{-\gamma}$, with $\gamma = \gamma_{\mathrm{SN}} -2$. However, only eleven star-forming galaxies are identified as gamma-ray sources at this time, including three LIRGs and one ULIRG, all four in GOALS \cite{gammaraySFG}. It is therefore not possible to constrain $\gamma_{\mathrm{SN}}$ systematically for individual galaxies in the GOALS sample.

\subsubsection{Maximum cosmic-ray momentum}

The maximum proton energy reached in the acceleration process ($E_{\mathrm{max}}=p_{\mathrm{max}}c$) determines up to which energy neutrinos are significantly produced. About 5 $\%$ of the primary proton energy is transferred to the high-energy neutrino in an inelastic collision. As such, to produce neutrinos of $\sim$1 PeV, as observed with IceCube, a cosmic accelerator should be able to accelerate particles up to $E_{\mathrm{max}} \sim 100$ PeV. This reduces to $E_{\mathrm{max}} \sim$ 1$-$10 PeV to produce neutrinos of $\sim$100 TeV. 

Observed cosmic rays with energies up to $\sim$3 PeV are generally attributed to galactic supernovae. This is based on energy considerations and GeV-TeV gamma-ray observations of supernova remnants \cite{origincr, SNpevatron2}. Modeling efforts are also in favor of particles reaching energies of $\sim$10$-$100 PeV in supernova acceleration. This relies on the presence of sufficiently strong magnetic fields and/or the presence of a magnetic-plasma wind of the progenitor star \cite{SNpevatron, SNpevatron3,SNpevatron4,SNpevatron5}. In (circum-)nuclear starburst regions of U/LIRGs, the magnetic field strength is significantly amplified (e.g.$~$\cite{Bmin}), and the newly formed stars could be on average more massive as opposed to normal star-forming regions. The latter implies an increase in the average explosion energy per supernova and an enhancement in the stellar-mass loss via stellar winds. This, in combination with the amplified magnetic field, indicates on average larger $p_{\mathrm{max}}$ values as opposed to star-forming galaxies such as the Milky Way.

\subsection{Cosmic-ray propagation and calorimetric conditions}

To model the confinement of the supernova-injected particles in the nuclear disk, a leaky-box model is assumed. This model dictates that the injected cosmic rays are allowed to move freely in the starburst volume and have a nonzero chance to escape the boundaries. The rate at which particles escape these boundaries is defined as the inverse of the average escape time $\tau_{\mathrm{esc}}$. We consider advection via a galactic-scale outflow and spatial diffusion as cosmic-ray removing processes. The average escape time $\tau_{\mathrm{esc}}$ in a particular starburst region is thus computed as
\begin{equation}
    \tau_{\mathrm{esc}} = \left[ \tau^{-1}_{\mathrm{diff}}+ \tau^{-1}_{\mathrm{adv}} \right]^{-1}~,
    \label{escapetime}
\end{equation}

\noindent with $\tau_{\mathrm{diff}}$ and $\tau_{\mathrm{adv}}$ the average timescales over which diffusion and advection occur, respectively. In addition, cosmic rays can also participate in inelastic pp-interactions before being removed from the starburst volume. These catastrophic collisions result in energy loss over an average timescale $\tau_{\mathrm{pp}}$. Continuous energy losses such as Coulomb interactions and ionization also affect the propagation of cosmic rays. However, for particle energies larger than 1 GeV, such energy losses are negligible as opposed to the catastrophic interactions (e.g.$~$\cite{Schl94}). Therefore, the continuous energy-loss processes can be safely neglected for the purposes of this work as a primary cosmic-ray energy of $E \gtrsim  1$ PeV is required to produce neutrinos at the level of IceCube observations. Moreover, for typical magnetic-field strengths at the scale of the starburst region, proton-synchrotron losses are negligible.

The average total time a particle spends in the starburst region $\tau$ is then computed as 
\begin{equation}
    \tau = \left[\tau^{-1}_{\mathrm{diff}} + \tau^{-1}_{\mathrm{adv}} + \tau^{-1}_{\mathrm{pp}}\right]^{-1}~.
    \label{restime}
\end{equation}

\noindent The diffusion timescale, advection timescale, and the energy-loss timescale due to inelastic pp-collisions are discussed in more detail in the following sections.

\subsubsection{\label{subsec: diff} Diffusion}

Cosmic rays injected by supernova activity will interact with the turbulent magnetic field in the starburst region. This leads to a random walk driven by the Larmor radius $r_{\mathrm{L}}$ of the particle. Eventually, the random walk leads to diffusion from the central starburst region, assuming no other processes are affecting the propagation. The timescale over which diffusion happens is therefore conservatively approximated as 
\begin{equation}
    \tau_{\mathrm{diff}} = \frac{H^2_{\mathrm{SBR}}}{D}~,
    \label{diffusion}
\end{equation}

\noindent with $D$ the diffusion coefficient, which depends on the magnetic field strength $B$ in the central starburst region, and $H_{\rm SBR}$ the scale height of the nuclear disk introduced in Sec.~\ref{geometry}.

Following \cite{Peretti1} we choose a Kolmogorov-type diffusion in the starburst volume. As such, the diffusion coefficient in Eq.$~$(\ref{diffusion}) is parametrized as 

\begin{equation}
    D = \frac{r_L c}{3 \mathcal{F}(k)}~,
\end{equation}

\noindent which is based on the quasi-linear formalism. The value of the diffusion coefficient $D$ scales with the relativistic gyroradius $r_\mathrm{L} = p/qB$ of the cosmic ray, with $q$ the charge, $p$ the momentum, and $B$ magnetic field strength in the central starburst region. The strength of the magnetic field in the central region of a starburst galaxy is typically $\gtrsim 100$ $\mu \mathrm{G}$ and can even reach a few mG \cite{Bmin}. For all U/LIRGs in this work, we fix $B = 250$ $\mu \mathrm{G}$, consistent with targeted observations of NGC 3690 (see Sec.~\ref{casestudy}). Furthermore, the diffusion coefficient is also affected by the speed of the cosmic ray, which is fixed to the speed of light $c$. $\mathcal{F}(k)$ is the normalized energy density  per unit logarithmic wave number $k$ in the turbulent magnetic field. This parameter is expressed as $\mathcal{F}(k) = k \cdot W(k) = k \cdot W_0 \cdot (k/k_0)^{-d}$ and normalized as
\begin{equation}
    \int^{\infty}_{k_0} \mathcal{F}(k) \mathrm{d}(\ln k) = \int^{\infty}_{k_0}  W_0 \cdot \left(\frac{k}{k_0}\right)^{-d} \mathrm{d}k = \eta_B~.
\end{equation}

\noindent Here $\eta_{B}$ = ($\delta B/B)^2$ is the turbulence ratio with $\delta B$ the turbulent component of the magnetic field, and $k^{-1}_0 = 1$ pc is the characteristic length scale at which turbulence is injected. In this work, we consider cosmic-ray interactions with large-scale Kolmogorov turbulence such that $d$ = 5/3 and $\eta_B = 1$. 

To evaluate the diffusion timescale $\tau_{\mathrm{diff}}$, it is assumed that cosmic rays predominantly interact with the resonant mode $k_{\mathrm{res}} = 1/r_L$. Then, $\mathcal{F}(k_{\mathrm{res}}) \propto k_{\mathrm{res}}^{-\frac{2}{3}} = r^{\frac{2}{3}}_L \propto  p^{\frac{2}{3}}$. As a result, the diffusion coefficient scales with momentum as $D(p) \propto p^{\frac{1}{3}}$ such that $\tau_{\mathrm{diff}} \propto p^{-\frac{1}{3}}$.

\subsubsection{\label{subsec: advection} Advection}

Galactic-scale outflows in starburst galaxies are commonly observed (see e.g.$~$\cite{VeilleuxREV} for a review). A possible driving mechanism for such an outflow is the mechanical energy transfer to the nuclear ISM via stellar winds and supernova explosions. These interactions induce strong shocks that heat and pressurize the ISM. In addition, AGN activity and cosmic rays are also proposed as driving mechanisms (e.g.$~$\cite{VeilleuxREV}). As a result of the energy transfer to the ISM, a cavity of very hot gas is formed. Due to the pressure imbalance between the nuclear region and the ISM of the host galaxy, this gas starts expanding above and below the galactic disk. Once the scale height of the galactic disk is reached, the wind breaks out into the galactic halo \cite{CC85} and thereby advects part of the cosmic-ray population out of the nuclear region. As such, these cosmic rays will not contribute to the high-energy neutrino production in the nuclear region. It should be noted that advected cosmic rays could be accelerated and converted to high-energy neutrinos within the wind \cite{Windneutrinos}. This contribution is not considered within our framework.

The velocity profile of the expanding bubble is such that the wind speed increases as the edge of the nuclear region is reached. As the expanding wind breaks out into the galactic halo, the terminal velocity ($v_{\infty}$) is quickly reached \cite{CC85}. The wind speed at the point of cosmic-ray advection ($v_{\mathrm{adv}}$) is therefore bound by the terminal velocity, i.e.$~v_{\mathrm{adv}} < v_{\infty}$.

The velocity of galactic-scale outflows is inferred from spectral line emission of the wind. Strong winds with speeds of 500$-$1500 km s$^{-1}$ have been detected by \textit{Herschel} in ULIRGs \cite{Herscheloutflow}. These winds are also observed in LIRGs. ALMA observations of the LIRG NGC 3256, for example, reveal a molecular outflow from the northern nuclear disk. This outflow is part of a starburst-driven superwind with a maximum velocity $>$ 750 km s$^{-1}$ \cite{2014ApJ...797...90S}. As indicated above, the advection speed is likely smaller than these terminal velocities.

The advection timescale $\tau_{\mathrm{adv}}$ is approximated as the ratio of the scale height of the nuclear disk $H_{\mathrm{SBR}}$ and the advection speed $v_\mathrm{adv}$,
 \begin{align}
     \tau_{\mathrm{adv}} = \frac{H_{\mathrm{SBR}}}{v_{\mathrm{adv}}}~.
 \end{align}

\subsubsection{\label{subsec: inelastic collisions} Energy-loss timescale }

To evaluate the rate at which cosmic rays lose their energy by inelastically colliding with the nuclear ISM, we make the assumption that the cosmic rays encounter the average ISM proton density in the nuclear region ($n$). The rate at which high-energy protons interact in the starburst region via inelastic pp-collisions then scales with the average ISM proton density $n$, the cross section of inelastic pp-collisons ($\sigma_{\mathrm{pp}}$), and the velocity of the cosmic ray. As the cosmic-ray protons of interest are highly relativistic, the speed of the cosmic rays is fixed to the speed of light $c$. The inelasticity of a collision is fixed to $\zeta = 0.5$ \cite{Inelas}. As such, the timescale for energy loss via inelastic pp-scattering can be expressed as
 
\begin{align}
    \tau_{\mathrm{pp}} = \frac{1}{n \cdot\sigma_{\mathrm{pp}}(E)\cdot c \cdot \zeta}.
    \label{collisiontsc}
\end{align}
     
\noindent For the cross section, we use the parametrization given in \cite{Kelner2006}, constructed from accelerator and simulation data, such that $\sigma_{\mathrm{pp}} =34.3+1.88 \ln L+0.25L^{2}~\mathrm{mb}$ with $L = \ln(E/1~\mbox{TeV})$.

\subsubsection{\label{subsec: calorimete}Calorimeter conditions}

A starburst region efficiently converts high-energy protons into neutrinos if the energy-loss timescale is significantly shorter than the timescale over which diffusion and advection occur. In that case, the starburst region acts as a calorimeter. To quantify the calorimeter conditions, the parameter $\mathcal{C}_{\mathrm{pp}} \in [0,1]$ is introduced as
\begin{equation}
    \mathcal{C}_{\mathrm{pp}} = \frac{\tau}{\tau_{\mathrm{pp}}} = \frac{f_{\mathrm{pp}}}{1+f_\mathrm{pp}}~.
\end{equation}

\noindent The parameter $f_{\mathrm{pp}}$ 
is the effective optical depth for pp-interactions, also known as the pp-collision efficiency, and is defined as the ratio of $\tau_{\mathrm{esc}}$ to $\tau_\mathrm{pp}$. As such, if the pp efficiency is large, secondary particle production will dominate over particle escape, which corresponds to $\mathcal{C}_{\mathrm{pp}}$ $\to$ 1. Conversely, if particle escape dominates, then $\mathcal{C}_{\mathrm{pp}} \ll 1$. Between these two extremes, $\mathcal{C_{\mathrm{pp}}} > 0.5$ indicates the conditions for which $\tau_{\mathrm{pp}}$ is on average the shortest timescale in the system. 

\begin{figure}
    \centering
    \includegraphics[width=0.50\textwidth]{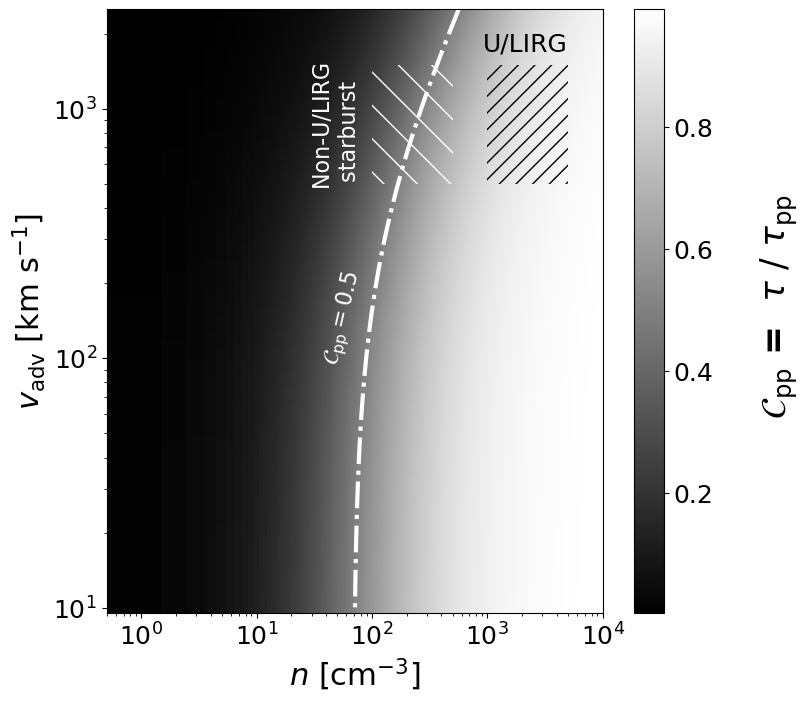}
    \caption{Parameter space of $\mathcal{C}_{\mathrm{pp}}$ for variable ISM proton density in the nuclear region ($n$) and advection speed ($v_{\mathrm{adv}}$). A 10 PeV proton is assumed to propagate in a nuclear disk with scale height $H_{\mathrm{SBR}} = 150$ pc and the diffusion is fixed to a Kolmogorov-type diffusion model. The free parameters of the diffusion model are fixed as discussed in Sec.~\ref{subsec: diff}. The white dash-dotted line shows the combinations of $v_{\mathrm{adv}}$ and $n$ for which $\mathcal{C}_{\mathrm{pp}}$ = 0.5. The white and black hatched regions indicate expected $\mathcal{C}_{\mathrm{pp}}$ values for nuclear starburst regions in non-U/LIRGs and U/LIRGs, respectively.}
    \label{heatmap}
\end{figure}

Figure \ref{heatmap} shows the parameter space of $\mathcal{C}_{\mathrm{pp}}$ for variable ISM proton density in the nuclear region ($n$) and advection speed ($v_{\mathrm{adv}}$). For this, a 10 PeV proton is assumed to propagate in a nuclear disk with scale height $H_{\mathrm{SBR}}$ = 150 pc, taking a Kolmogorov-type diffusion model. The dash-dotted line shows for which combinations of $n$ and $v_{\mathrm{adv}}$ a value of $\mathcal{C_{\mathrm{pp}}}$ = 0.5 is obtained. The black hatched region indicates $\mathcal{C}_{\mathrm{pp}}$ values for typical ISM proton densities in the nuclear region of U/LIRGs, $n$  $\gtrsim 1000$ cm$
^{-3}$ (\cite{Nucleardiskdensity}, see also Sec.~\ref{casestudy}), and advection speeds between 500 km s$^{-1}$ and 1500 km s$^{-1}$. Note that, although terminal velocities of $\sim$1500 km s$^{-1}$ are observed in U/LIRGs, it is unlikely that the advection speed $v_{\mathrm{adv}}$ is equally high (Sec.~\ref{subsec: advection}). 

In \cite{Spitzer3}, for example, it is shown that GOALS U/LIRGs in a late or final merger stage are on average more obscured. As galaxies merge, gas and dust are funneled toward the central regions, making them more compact and obscured. As such, ULIRGs, which are nearly always in the final stage of a merger, are expected to be located at the high end of the particle densities indicated in Fig.~\ref{heatmap}. For LIRGs, which are observed in every merger stage, this could strongly depend on how advanced the merger is. In any case, high-energy protons are expected to lose a significant fraction of their initial energy in the nuclear region of U/LIRGs. For comparison, we also investigate typical conditions in non-U/LIRG starburst galaxies. Prototypical examples of such galaxies are the nearby starburst galaxies M82 and NGC 253. This type of starburst galaxy typically has a lower ISM proton density in the nuclear region, i.e.~$n\sim 100$ cm$^{-3}$ (e.g.$~$\cite{NGC253, Peretti1, M82YOAST,radiogammacorr}). The white hatched region in Fig.~\ref{heatmap} shows $C_{\mathrm{pp}}$ values corresponding to ISM proton number densities between 100 and 500 cm$^{-3}$, and the same advection speeds as investigated for the U/LIRGs. Compared to U/LIRGs, non-U/LIRG starburst galaxies are expected to be less efficient calorimeters on average. 

\begin{figure}
    \centering
    \includegraphics[width=0.44\textwidth]{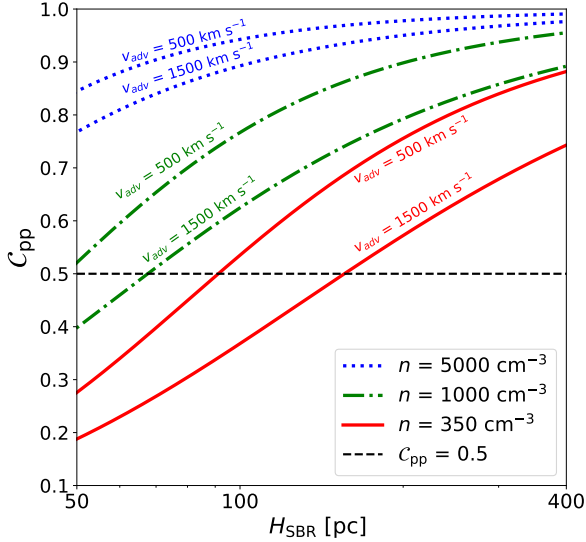}
    \caption{The $\mathcal{C}_{\mathrm{pp}}$-parameter for three different starburst configurations at variable scale height ($H_{\mathrm{SBR}}$). The ISM proton density is fixed to $n$ = 350 cm$^{-3}$, $n$ = 1000 cm$^{-3}$, and  $n$ = 5000 cm$^{-3}$. For each of these configurations, a galactic superwind with $v_{\mathrm{adv}}$ = 500 km s$^{-1} $ and $v_{\mathrm{adv}} = 1500$ km s$^{-1}$ is considered.}
    \label{H}
\end{figure}

The scale height of the nuclear disk ($H_{\mathrm{SBR}}$) also affects $\mathcal{C}_{\mathrm{pp}}$. Figure \ref{H} shows how $\mathcal{C}_{\mathrm{pp}}$ is affected when varying $H_{\mathrm{SBR}}$ between 50 and 400 pc for a starburst region with a nuclear ISM density of $n$ = 350 cm$^{-3}$, $n$ = 1000 cm$^{-3}$, and  $n$ = 5000 cm$^{-3}$. The range of scale heights is consistent with the values derived for nearby U/LIRGs \cite{2014ApJ...784...70M}. The ISM particle density values are chosen to model a wide range of starburst conditions. For each of these starburst configurations, an advection speed of $v_{\mathrm{adv}}$ = 500 km s$^{-1} $ and $v_{\mathrm{adv}} = 1500$ km s$^{-1}$ is considered. The results show that calorimeter assumptions are robust against changes in $H_{\mathrm{SBR}}$ and $v_{\mathrm{adv}}$ if the particle density in the nuclear region is high. This statement also applies to changes in the diffusion model. 

\subsection{From cosmic-ray injection to neutrino production at the source}

The distribution of high-energy proton momenta in the nuclear region of U/LIRGs ($\mathcal{F}_\mathrm{p}$) is determined by the interplay between the injection rate of high-energy protons by supernovae and subsequent particle transport, as described above. Assuming a spatially homogeneous starburst region in a steady state, the momentum distribution of high-energy protons in the nuclear region is expressed as
\begin{equation}
    \mathcal{F}_\mathrm{p} = Q_\mathrm{p} \cdot \tau = Q_ \mathrm{p}\cdot \tau_{\mathrm{pp}} \cdot \mathcal{C}_{\mathrm{pp}} ~. 
\end{equation}

High-energy protons can collide inelastically with a proton in the nuclear ISM. Such collisions produce, among other particles, charged ($\pi^{\pm}$) and neutral pions ($\pi^0$). The charged pions decay as

\begin{gather}
\begin{cases}
\pi^+ \to \mu^+ + \nu^{(1)}_\mu \to e^+ + \nu_e + \bar{\nu}^{(2)}_\mu + \nu^{(1)}_\mu\\
\pi^- \to \mu^-+\bar{\nu}^{(1)}_\mu \to e^- + \bar{\nu}_e + \nu^{(2)}_\mu + \bar{\nu}^{(1)}_\mu~,
\label{chargpiondecay}
\end{cases}
\end{gather}

\noindent and the neutral pions decay to gamma rays, $\pi^0 \to \gamma\gamma$. The label (1) denotes the muon neutrinos produced in pion decay and the label (2) denotes the muon neutrinos produced in muon decay. To compute the neutrino production rate ($q_\nu$) from the energy distribution of high-energy protons, $n_\mathrm{p}(E) = 4\pi p^2 \mathcal{F}_\mathrm{p}(p)\mathrm{d}p$, we follow the approach outlined in \cite{Kelner2006}. The authors provide analytical fits to neutrino spectra obtained from meson spectra simulated with Monte Carlo generators SYBILL and QGSJET. Doing so, the neutrino production rate $q_{\nu}$ at the source, including neutrinos and antineutrinos, is expressed as 

\begin{equation}
    q_{\nu} = c n \int^{1}_{0} F_{\nu}\left(x,\frac{E_{\nu}}{x}\right)\sigma_{\mathrm{pp}}\left(\frac{E_\nu}{x}\right)n_\mathrm{p}\left(\frac{E_\nu}{x}\right)\frac{\mathrm{d}x}{x}~,
\end{equation}

\noindent where $x = E_\nu/E_{\mathrm{p}}$ and $F_{\nu}$ = $F^{(1)}_{\nu_\mu} + F^{(2)}_{\nu_\mu} + F_{\nu_e}$ are the neutrino distribution functions corresponding to the decays given in Eq.~(\ref{chargpiondecay}). The spectrum of the muon neutrinos and electron neutrinos obtained from muon decay are described by $F^{(2)}_{\nu_\mu}$ and $F_{\nu_e}$, respectively. The former is described by the same function that describes electrons produced in muon decay, $F_e$. Moreover, $F_{\nu_e} \approx F_e$ within 5 $\%$. As such, we use  $F_{\nu} = $  $F^{(1)}_{\nu_\mu} + 2\cdot F_{e}$. The distribution functions for $F_e$ and for muon neutrinos produced in pion decay, $F^{(1)}_{\nu_\mu}$, correspond to Eq.$~$(62)$-$(65) and Eq.$~$(66)$-$(69) in \cite{Kelner2006}, respectively. Note that these analytical fits can only be used for secondaries with energies larger than 100 GeV. 

Integrating the neutrino-production rate over the volume of the starburst region yields the neutrino luminosity. Therefore, the all-flavor neutrino flux at Earth from a single GOALS galaxy ($\Phi_{\nu}$), containing neutrinos and antineutrinos, is computed as
\begin{equation}
    \Phi_{\nu}(E,z) = \frac{V_{\mathrm{SBR}}}{4\pi D^2_L} \cdot q_{\nu}(E(1+z))~,
    \label{totalflux}
\end{equation}

\noindent with $D_L$ the luminosity distance to the galaxy and $z$ its redshift. Note that $\Phi_\nu \propto E^{-\gamma}$ with $\gamma = \gamma_{\mathrm{SN}}-2$.

Inelastic pp-interactions at the source result in a neutrino-flavor ratio given by $(\nu_e : \nu_\mu : \nu_\tau)=(1 : 2 : 0)$. However, the combination of propagating over extragalactic distances and neutrino oscillations leads to an approximately equal distribution among the three neutrino flavors. As such, the flavor ratio at Earth is expected to be $(\nu_e : \nu_\mu : \nu_\tau) \approx (1 : 1 : 1)$ \cite{2006MPLA...21.1049A}. This implies that the single-flavor neutrino flux at Earth ($\Phi_{\nu_j}$), with $j \in \{e,\mu,\tau\}$, is obtained by dividing the all-flavor neutrino flux by a factor three. The muon-neutrino flux is of particular interest in the search for the origin of astrophysical neutrinos observed with IceCube, as discussed in Sec.~\ref{sec:level1}.

In conclusion, within the framework considered in this study, the neutrino flux depends on the starburst-specific parameters
\begin{equation}
     \Phi_{\nu_j} = \Phi_{\nu_j}\left( \mathcal{R}_{\mathrm{SN}}, \gamma_{\mathrm{SN}}, p_{\mathrm{max}} , H_{\mathrm{SBR}}, v_{\mathrm{adv}}, n, B,  D_L\right),
    \label{endresult}
\end{equation}

\noindent for a particular diffusion model in the nuclear region and the supernova rate computed as $\mathcal{R}_{\mathrm{SN}} = \mathcal{R}_{\mathrm{SN}}\left(L_{\mathrm{IR}}, \langle \alpha_\mathrm{AGN} \rangle, \mathcal{G}\right)$.

\section{Case study: LIRG NGC 3690 \label{casestudy}}

\begin{figure}
    \centering
    \includegraphics[width=0.35\textwidth]{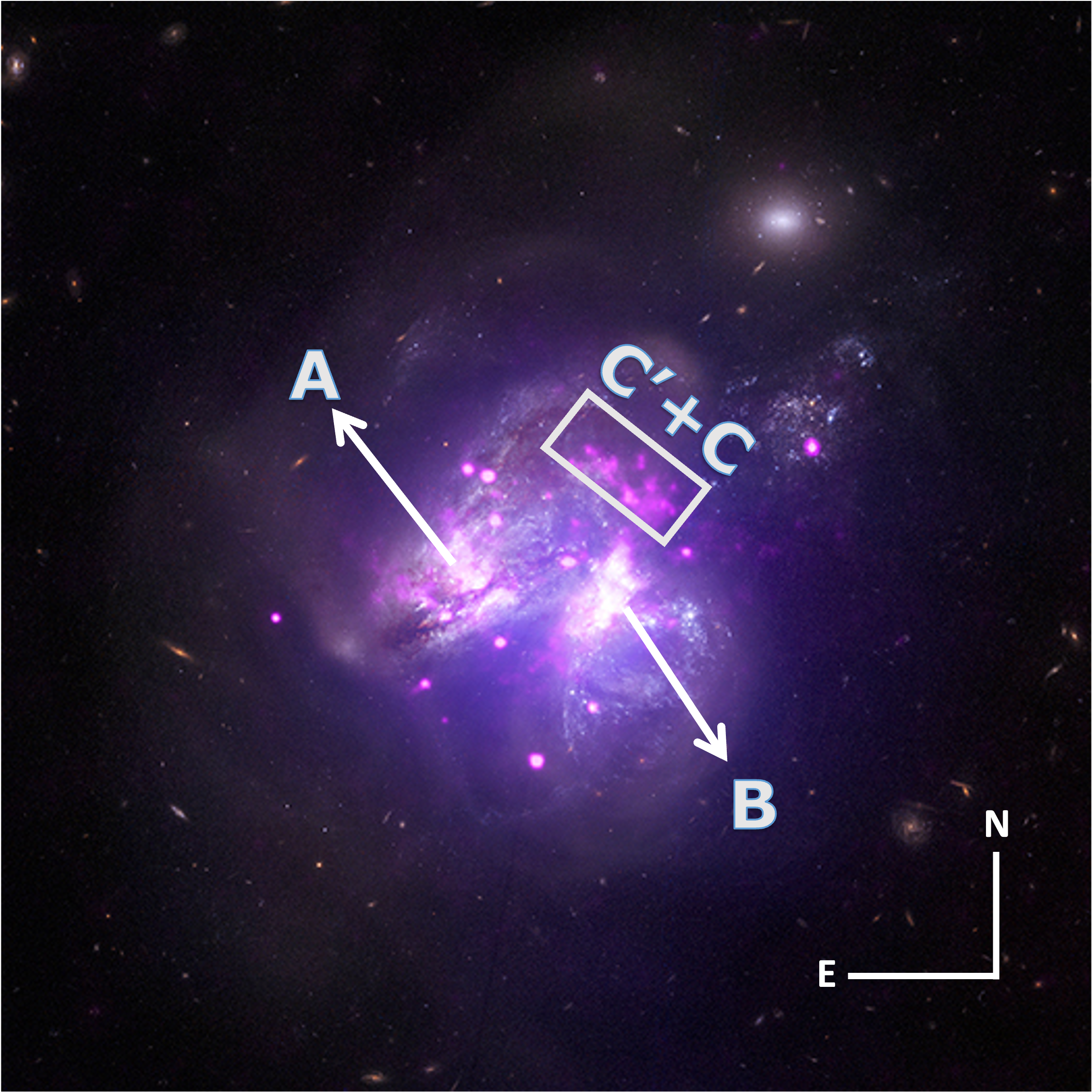}
    \caption{Composite image of NGC 3690 showing both optical (white) and x-ray (pink) emission. The two nuclear regions A (east) and B (west) are indicated, as well as the off-nuclear region C'+C. Edited from \cite{WinNT}.}
    \label{Arp299Figure}
\end{figure}

In this section, the starburst-driven neutrino-production framework is applied to the LIRG NGC 3690 (also known as Arp 299 and Mrk 171)\footnote{In the literature, the eastern and western members of the galaxy pair in Fig.~\ref{Arp299Figure} are often given the names IC 694 and NGC 3690, respectively. However, IC 694 is actually a small E/S0 galaxy $\sim$1' to the northwest of the merging galaxy pair, while NGC 3690 properly refers to the merging pair \cite{1999ApJ...517..130H}. The merger is also commonly known as Arp 299. In this work, we choose the name NGC 3690 to refer to the whole system, consistent with the name given by GOALS.}. This intermediate-stage merger between two gas-rich galaxies, shown in Fig.~\ref{Arp299Figure}, is one of the most powerful merging galaxies in the local Universe at a luminosity distance $D_L \sim 50.7$ Mpc \cite{GOALS}. It is located in the Northern Hemisphere at equatorial coordinates $\alpha_{\mathrm{J2000}}$ = 11h28m32.3s and $\delta_{\mathrm{J2000}}$ = 58d33m43s. The eastern part of the galaxy  system (NGC 3690E) has a \textit{Herschel} luminosity of log$_{10}$($L_{\mathrm{IR}}$/$L_{\odot}$) = 11.37 and the western part (NGC 3690W) has a \textit{Herschel} luminosity of log$_{10}$($L_{\mathrm{IR}}$/$L_{\odot}$) = 11.09 \cite{HerschelSantos}. Mid-IR and radio continuum maps of this LIRG reveal distinct regions A, B, and C+C' which dominate at these wavelengths \cite{arp299regions}. Region A is the nuclear region of the eastern galaxy (NGC 3690E-A) and region B is the nuclear region of the western part (NGC 3690W-B). The C'+C component is located in the overlapping region between the two galaxies. Multi-wavelength follow-up studies show that the nature of the nuclear regions are very different. Hard x-ray observations indicate the presence of a Compton thick AGN in region B \cite{Arp299AGN} and high-resolution radio observations reveal a strong nuclear starburst in region A \cite{2012A&A...539A.134B}.  We also note that a tidal disruption event (TDE) was observed in region B \cite{Arp299tde}. 

Region A is of main interest for this work and is used as a case study for the neutrino production model introduced in Sec.~\ref{framework}. In the following, we first identify the parameters related to cosmic-ray injection in region A, followed by those related to cosmic-ray propagation. Finally, we use these parameters to estimate the starburst-driven muon-neutrino flux from region A in NGC 3690.

\subsubsection{Cosmic-ray injection}

NGC 3690E-A is characterized as a highly dust-enshrouded region, such that even near-IR wavelengths suffer from attenuation effects \cite{2012A&A...539A.134B}. Therefore, high-resolution radio observations are required to identify the supernova activity in this region. The best direct observational constraints on the supernova activity in the central $R_{\mathrm{SBR}}$ = 150 pc of NGC 3690E-A were revealed by a $\sim$2.5 year monitoring campaign at 5.0 GHz. This campaign revealed two CCSN in the starburst region leading to an estimated lower limit of $\mathcal{R}_{\mathrm{SN}} \gtrsim 0.80^{+1.06}_{-0.52}$ yr$^{-1}$ with uncertainties corresponding to 1$\sigma$ errors \cite{2012A&A...539A.134B}. Besides these direct observations, the authors also present a supernova rate estimated from diffuse synchrotron observations that is found to be $\mathcal{R}_{\rm SN} \simeq 0.45-0.65$ yr$^{-1}$. This estimate agrees well with our supernova rate estimates computed via Eq.~(\ref{RSNformula}), given in Table \ref{tab: cosmicraylum}. This table also provides the corresponding mass-scaling factor $\mathcal{M}$ (Sec.~\ref{M}) and cosmic-ray luminosity $\mathcal{L}_{\mathrm{CR}}$. The latter is the relevant parameter to compute the neutrino flux. To convert the supernova rates to a cosmic-ray luminosity, the  kinetic energy conversion factor is fixed to $\eta_{\mathrm{SN}} = 0.10$.

The spectral index of the proton injection rate due to supernova activity $\gamma_{\mathrm{SN}}$ is determined from the spectral index of the gamma-ray spectrum $\Gamma$. This means, $\gamma_{\mathrm{SN}} =  \Gamma_{\mathrm{NGC3690}} + 2$ with $\Gamma_{\mathrm{NGC3690}}$ = 2.11 $\pm$ 0.19 determined from gamma-ray observations \cite{gammaraySFG}. It is noted that the location of the gamma-ray emission in NGC 3690 is unresolved. As such, the gamma rays could also (partially) originate from the AGN in region B or an off-nuclear star-forming region. To demonstrate the effect of changing the spectral index, the neutrino flux is also computed for  $\gamma_{\mathrm{SN}}$ = 4, which corresponds to $\Gamma = \gamma = 2$. 

The maximum proton momentum $p_{\mathrm{max}}$ for the supernova activity in NGC 3690E is unconstrained by data. Therefore, we investigate the neutrino flux for a maximum momentum $p_{\mathrm{max}}$ of 10 PeV/$c$, 20 PeV/c, 30 PeV/$c$, and 100 PeV/$c$.

\begin{table}
\caption{\label{tab: cosmicraylum}
 The supernova rate $\mathcal{R}_{\mathrm{SN}}$ of the nuclear region in NGC 3690E for different IMFs together with the progenitor mass-scaling factor $\mathcal{M}$, and the cosmic-ray luminosity $\mathcal{L}_{\mathrm{CR}}$. The supernova rate is computed using Eq.~(\ref{RSNformula}) with log$_{10}$($L_{\mathrm{IR}}$/$L_{\odot}$) = 11.37, $\langle \alpha_{\mathrm{AGN}} \rangle$ = 0.04 (see Sec.~\ref{alphagn}), and $\mathcal{G}$ = 1. To compute the cosmic-ray luminosity $\mathcal{L}_{\mathrm{CR}}$, we take $\eta_{\mathrm{SN}}$ = 0.10 in Eq.~(\ref{CRlumformula}).}
\begin{ruledtabular}
\begin{tabular}{ccccc}
 & $\mathcal{R}_{\mathrm{SN}}$ [yr$^{-1}$] & $\mathcal{M}$ & $\mathcal{L}_{\mathrm{CR}}$  [10$^{49}$~erg yr$^{-1}$]  \\
\hline
Salpeter & 0.52 & 0.93 & 4.85 & \\
Kroupa & 0.51 & 1.00 & 5.14   \\
TH  $\beta_{\mathrm{high}} = 1.5$ & 0.36 &   1.37 & 4.98  \\
TH $\beta_{\mathrm{high}} = 1.0$ & 0.28 & 2.01 & 5.59  \\
\end{tabular}
\end{ruledtabular}
\end{table}

\subsubsection{Cosmic-ray transport and calorimeter conditions}

Aperture synthesis CO maps of NGC 3690 show an H$_2$ mass of 3.9 $\times$ 10$^{9}$ M$_{\odot}$ in the central $\leq$ 250 pc of NGC 3690E \cite{1991ApJ...366L...1S}. Here, we assume that this mass is distributed in a uniform way within a disk with radius of $R_{\mathrm{SBR}}$ = 250 pc and scale height $H_{\mathrm{SBR}}$ = 150 pc. This corresponds to an average H$_2$ particle density of 

\begin{equation}
    n_{\mathrm{H}_2} \approx 1340 \cdot \left(\frac{H_{\mathrm{SBR}}}{150 \mathrm{pc}}\right)^{-1} \cdot \left(\frac{R_{\mathrm{SBR}}}{250 \mathrm{ pc}}\right)^{-2}~\mathrm{cm}^{-3}~.
\end{equation}

\noindent The interest is, however, in the proton number density $n$ which is a factor of two larger. It is noted that the proton density inferred from the H$_2$ number density is a lower limit on the proton density in the ISM as there is also a subdominant contribution of neutral atomic hydrogen (e.g.~\cite{neutralatomicH}) and heavier elements. For this work, therefore, we make the conservative choice of $n = 2500$ cm$^{-3}$ for the ISM proton number density.

Observations of NGC 3690E with the International \textit{Low Frequency Array} (LOFAR) Telescope at 150 MHz show a two-sided, wide filamentary structure emanating from the nucleus \cite{Arp299wind2018}. The outflow is detected via radio wavelengths from synchrotron-emitting electrons. Under the assumption that the outflow is driven by a supernova rate of $\mathcal{R}_{\mathrm{SN}} \gtrsim 0.80$ yr$^{-1}$, the outflow is estimated to move at 370-890 km s$^{-1}$ \cite{Arp299wind2018}. Here we assume an advection speed of $v_{\mathrm{adv}}$ = 500 km s$^{-1}$, consistent with observations of other U/LIRGs (Sec.~\ref{subsec: advection}). Moreover, LOFAR observations at 150 MHz indicate a minimum equipartition magnetic field for the nuclear region of $B \gtrsim 250$ $\mu \mathrm{G}$ \cite{2022A&A...658A...4R}. To compute the neutrino flux, a magnetic field strength of $B = 250$ $\mu \mathrm{G}$ and Kolmogorov-type diffusion model are used. 

\begin{figure}
    \centering
    \includegraphics[width=0.44\textwidth]{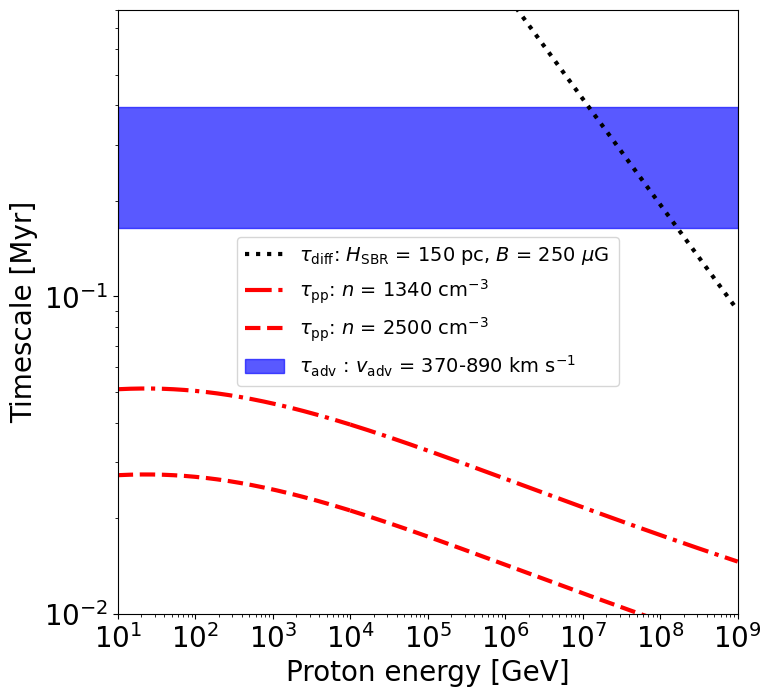}
    \caption{Diffusion, advection, and pp-energy loss timescales as function of proton energy for the nuclear starburst region of NGC 3690E\label{timescales}.}
\end{figure}

Figure \ref{timescales} shows the diffusion timescale, advection timescale, and pp energy-loss timescale as functions of proton energy for the nuclear region of NGC 3690E. Moreover, based on the values found for $n$, $H_{\mathrm{SBR}}$, $v_{\mathrm{adv}}$, and $B$, it follows that $\mathcal{C}_{\mathrm{pp}}$ = 0.95 (Sec.~\ref{subsec: calorimete}). This implies that the nuclear starburst region in NGC 3690E is expected to efficiently convert cosmic-ray energy into high-energy neutrinos.

\subsubsection{Neutrino flux predictions}

\begin{figure}
    \includegraphics[width=0.45\textwidth]{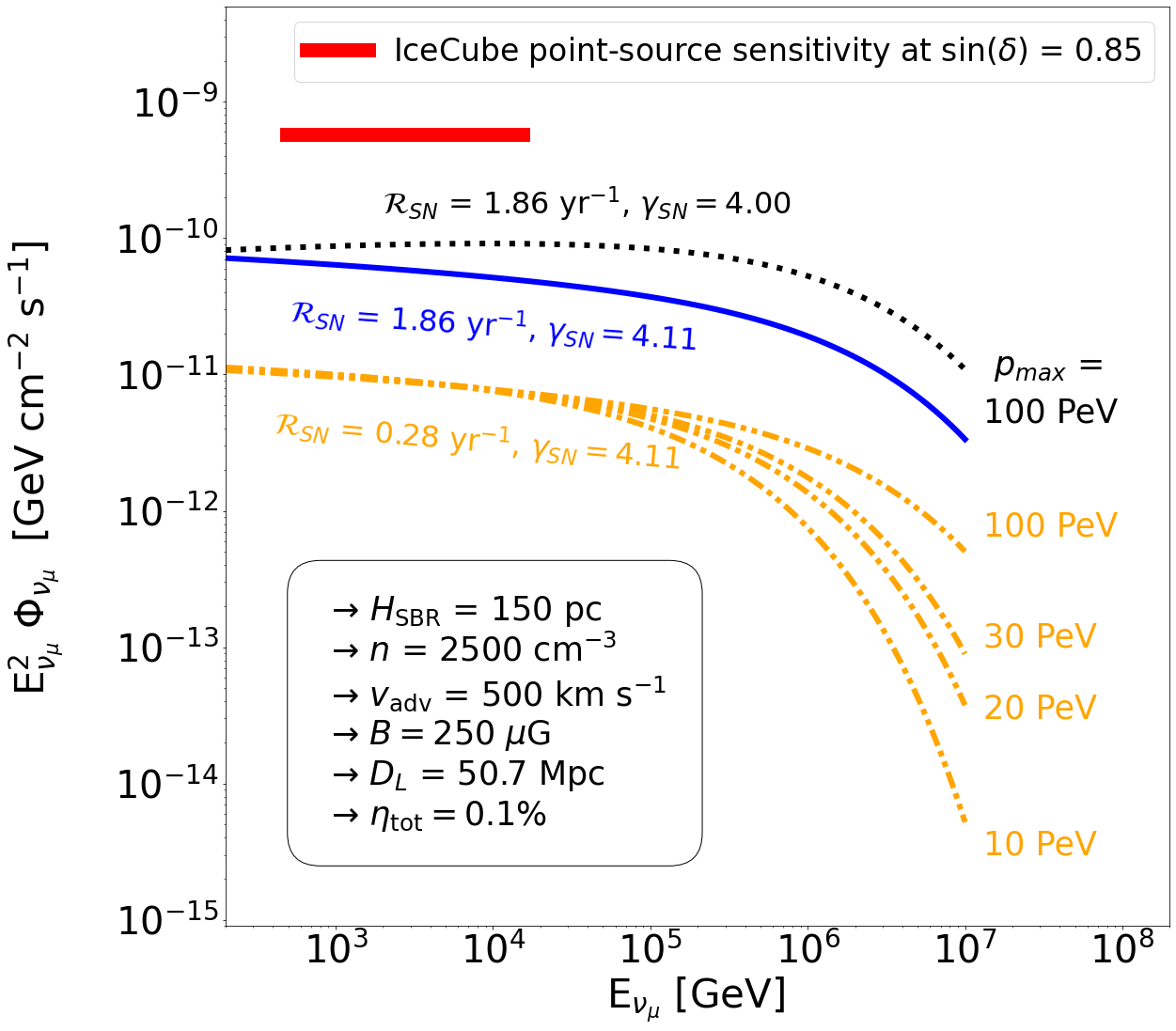}
    \caption{Predictions for the starburst-driven muon-neutrino flux of NGC 3690E using our model. All model parameters, except for the maximum proton momentum $p_{\mathrm{max}}$, are driven by multiwavelength observations as discussed in the text. Note that $\gamma = \gamma_{\mathrm{SN}} - 2$, where $\Phi_{\nu_\mu} \propto E^{-\gamma}$. The 10-year $E^{-2}$ IceCube point-source sensitivity for a source at the declination of NGC 3690 is also indicated by the red solid line \cite{2020PhRvL.124e1103A}.}
    \label{fluxarp299}
\end{figure}

The expected muon-neutrino flux for region A in NGC 3690E, for both a supernova rate of $\mathcal{R}_{\mathrm{SN}} = 0.28$ yr$^{-1}$ and $\mathcal{R}_{\mathrm{SN}}$ = 1.86 yr$^{-1}$, is shown in Fig.~\ref{fluxarp299}. The values of $\mathcal{R}_{\mathrm{SN}}$ correspond to the 1$\sigma$ errors on the direct observations. Figure \ref{fluxarp299} shows that the supernova rate affects the flux predictions linearly and that small changes in the spectral index $\gamma_{\mathrm{SN}}$ can have a significant effect on the flux predictions. It is noted that the high-energy tail of $E_{\nu_\mu}^2\Phi_{\nu_\mu}(E_{\nu_\mu})$ should be interpreted carefully. Although the exponential cutoff in the proton injection rate is a reasonable assumption, it is not driven by observations. Next-generation neutrino observatories, such as IceCube-Gen2 \cite{Gen2}, will help to test the validity of this exponential cutoff. The horizontal red line shows the point-source sensitivity based on 10 years of IceCube data for an $E^{-2}$ neutrino spectrum at the declination ($\delta$) of NGC 3690 \cite{2020PhRvL.124e1103A}. None of the investigated parameter combinations violate this sensitivity. This serves as a consistency check for the model since NGC 3690 has not shown up as a significant neutrino emitter in previous IceCube analyses.

\section{Diffuse flux predictions\label{diffuseflux}} 

In this section, first the per-source muon-neutrino flux generated by starburst activity ($\Phi_{\nu_\mu}$) is computed for the $N = 229$ GOALS galaxies targeted in this work (Sec.~\ref{sec1}). The calculations are done using our framework introduced in Sec.~\ref{framework}. Then, based on these predictions, the corresponding diffuse muon-neutrino flux ($\Phi^{\mathrm{diffuse}}_{\nu_\mu}$) from all GOALS galaxies is estimated as

\begin{equation}
    \Phi^{\mathrm{diffuse}}_{\nu_\mu}(E_{\nu_\mu}) = \frac{1}{4\pi}\sum^{N = 229}_{i = 1} \Phi_{i,\nu_\mu  }(E_{\nu_\mu})~.
    \label{diffuseGOALS}
\end{equation}

\noindent Finally, the diffuse flux from the total LIRG population integrated over cosmic history is estimated from a volume-limited sub-sample of GOALS.

\subsection{\label{PerandDiffuseGOALS} Per-source and diffuse neutrino flux estimates for the GOALS sample}

The eight parameters in Eq.$~(\ref{endresult})$ are required for each of the GOALS galaxies to compute their corresponding neutrino flux. However, the scale height ($H_{\mathrm{SBR}}$), the advection speed ($v_{\mathrm{adv}}$), the magnetic field strength ($B$), and the nuclear ISM proton density ($n$) are unknown for the majority of the GOALS galaxies. To fix these parameters, similar conditions are assumed as found in the case study of NGC 3690E (Sec.~\ref{casestudy}). The corresponding values are given in Table \ref{tab:protoU/LIRG}. In contrast to these fixed parameters, the supernova rate in the nuclear region ($\mathcal{R}_{\mathrm{SN}}$) is computed from source-specific data. The supernova rate per galaxy is computed with Eq.$~(\ref{RSNformula})$, using the \textit{Herschel} IR luminosity and the relative AGN contribution to the bolometric luminosity ($\langle \alpha_{\mathrm{AGN}} \rangle$). Both parameters are available for all 229 galaxies (see Sec.~\ref{sec1}). A canonical Kroupa IMF with $\Lambda_{\mathrm{IR}}$ = 5.97 $\times$ 10$^{-46}$ (Table \ref{tab:imf}) is used, and the assumption is made that half of the total IR luminosity of a galaxy is generated by the nuclear region, i.e.$~\mathcal{G} = 0.5$. We use these individual values to compute the diffuse neutrino flux expected from GOALS. In addition, we also compute the diffuse neutrino flux expected from GOALS for $\langle \alpha_{\mathrm{AGN}}\rangle$ = 0 in each source. This allows to investigate how an AGN contribution to the IR luminosity affects the expected neutrino flux.

To normalize the injection rate of high-energy particles to the nuclear supernova activity, one needs the spectral index of the injection spectrum of cosmic rays ($\gamma_{\mathrm{SN}}$), the maximum momentum reached in supernova acceleration ($p_{\mathrm{max}}$), and the conversion factor between supernova explosion energy and cosmic-ray acceleration ($\eta_{\mathrm{SN}}$). For each supernova event $\eta_{\mathrm{SN}}$ = 0.10 and $E_{\mathrm{SN}} = 10^{51}$ erg are assumed. However, the spectral index $\gamma_{\mathrm{SN}}$ is unconstrained for nearly all GOALS galaxies. Therefore, we opt to take the same spectral index in each galaxy and compute the diffuse neutrino flux for three different cases, i.e.$~\mathrm{for  }~\gamma_{\mathrm{SN}} = 4.00$, $\gamma_{\mathrm{SN}} = 4.25$, and $\gamma_{\mathrm{SN}} = 4.50$. This corresponds to $\gamma = \gamma_{\mathrm{SN}} - 2.00$, where $\Phi_{\nu} \propto E^{-\gamma}$. For all cases, an exponential cutoff in the proton injection spectrum is chosen at $p_{\mathrm{max}}$ = 100 PeV/$c$. Given all these parameters, the neutrino luminosity at the source can be computed for all 229 galaxies. To find the corresponding neutrino flux at Earth, the luminosity distances ($D_L$) provided by GOALS are used \cite{GOALS}. 

\begin{table}
\caption{\label{tab:protoU/LIRG}
Fixed parameters used in each of the GOALS galaxies to compute the per-source and diffuse muon-neutrino flux. }
\begin{ruledtabular}
\begin{tabular}{ccccc}
$p_{\mathrm{max}}$ [PeV/$c$]& $H_{\mathrm{SBR}}$ [pc] & $v_{\mathrm{adv}}$ [km s$^{-1}$] & $n$ [cm$^{-3}$] & $B$ [$\mu $G]  \\
\hline
100 & 150 & 500 & 1000  & 250 \\
\end{tabular}
\end{ruledtabular}
\end{table}

\begin{figure}
    \centering
\includegraphics[width=0.50\textwidth]{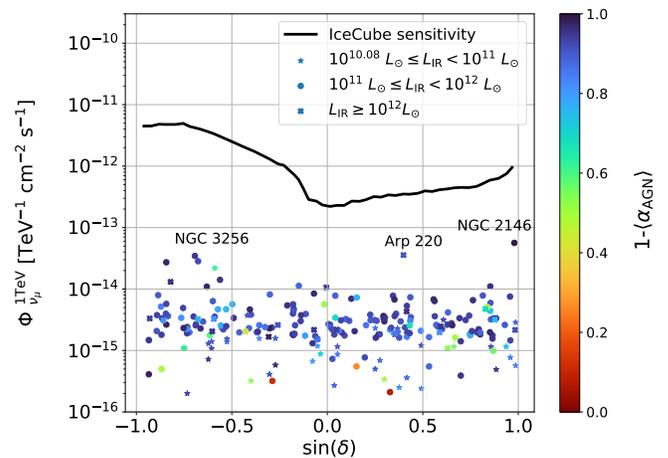}
   \caption{Per-source muon-neutrino flux predictions at 1 TeV as a function of the sine of the declination of the 229 GOALS galaxies targeted in this work. All fluxes are computed for a spectral index 
 $\gamma_{\mathrm{SN}}$ = 4, with $\gamma = \gamma_{\mathrm{SN}}-2$,  $\eta_{\mathrm{tot}} = 0.1 \%$, and the other model parameters as discussed in the text. The color scale indicates the value of $1-\langle \alpha_{\mathrm{AGN}} \rangle$ per galaxy. For each of the galaxies, it is indicated whether it is a galaxy with $10^{10.08} L_{\odot} \leq L_{\mathrm{IR}} < 10^{11} L_{\odot}$ (star), $10^{11} L_{\odot} \leq L_{\mathrm{IR}} < 10^{12} L_{\odot}$ (circle), or $L_{\odot} \geq 10^{12} L_{\odot}$ (cross). The 10-year $E^{-2}$ IceCube point-source sensitivity as function of the sine of the declination is also indicated \cite{2020PhRvL.124e1103A}.}
    \label{fig:persource}
\end{figure}

Figure \ref{fig:persource} shows the modeled per-source starburst-driven muon-neutrino fluxes at 1 TeV ($\Phi^{1\mathrm{TeV}}_{i,\nu_\mu}$) as a function of the sine of the declination of these sources. The fluxes are computed for $\gamma = 2$. For each galaxy, it is indicated whether it is a galaxy with $10^{10.08}L_{\odot} \leq L_{\mathrm{IR}} < 10^{11} L_{\odot}$, a LIRG, or an ULIRG. Moreover, the color scale indicates the $1-\langle \alpha_{\mathrm{AGN}} \rangle$ value of the corresponding galaxy. The plot also shows the 10-year $E^{-2}$ IceCube point source sensitivity at 1 TeV which is indicated by the solid black line. The three indicated galaxies in Fig.~\ref{fig:persource} are the top three galaxies with the strongest expected starburst-driven neutrino flux in the GOALS sample. It should be noted, however, that one of the most nearby GOALS galaxies, NGC 1068, is not shown in this plot. The $\langle \alpha_{\mathrm{AGN}} \rangle$-value of NGC 1068 is put to unity by GOALS. Therefore, this particular galaxy does not have a starburst-driven flux, as Eq.$~$(\ref{RSNformula}) indicates that the supernova rate would be zero. However, as NGC 1068 is so close to Earth, it follows that its $\langle \alpha_{\mathrm{AGN}} \rangle$-value is most likely smaller than unity. This is a result of the selection effect discussed in Sec.~\ref{sec1}. For $\langle \alpha_{\mathrm{AGN}} \rangle \leq 0.54$, NGC 1068 has the strongest expected neutrino flux out of all the investigated GOALS galaxies. This is a result of the proximity of the source and its moderate IR luminosity of $\log_{10}(L_{\mathrm{IR}}/L_{\odot})$ = 11.39 \cite{HerschelSantos}. In the most optimistic case, for $\gamma = 2$ and $\langle \alpha_{\mathrm{AGN}} \rangle$ = 0, our prediction for the starburst-driven muon-neutrino flux from NGC 1068 at 1 TeV is $\Phi^{1\mathrm{TeV}}_{\nu_\mu} = 2.44 \times 10^{-13}$ TeV$^{-1}$ cm$^{-2}$ s$^{-1}$. This prediction is about two orders of magnitude smaller than the neutrino flux from the direction of NGC 1068 reported by IceCube (see Sec.~\ref{sec:level1}). Note that the latter is compatible with a significantly softer spectral index $\gamma \approx 3.2$ \cite{Evidence1068}. Assuming that the neutrino flux is indeed generated by NGC 1068, our model suggests a dominant contribution from a nonstarburst component, such as AGN-related activity. Interestingly, the spectral index of the neutrino spectrum ($\gamma = 3.2 \pm 0.2$) is much softer than the spectral index of the gamma-ray spectrum ($\Gamma = 2.27 \pm 0.09$ \cite{gammaraySFG}). This hints toward different underlying processes giving rise to the observed neutrinos and the observed gamma rays. A possible scenario is that the observed gamma rays are produced by starburst activity in the circumnuclear starburst region (e.g.~\cite{1068puzzle}), while the observed neutrinos are produced in a gamma-ray opaque region close to the supermassive black hole of NGC 1068 (e.g.~\cite{hiddenhearth}). In that case the GeV-TeV gamma rays, produced simultaneously with the observed neutrinos, will not be detected as they cascade down to MeV energies or lower, which is below the detection threshold. 

Figure \ref{fig:persource} also illustrates that the most luminous IR sources in the GOALS sample, the ULIRGs, are not necessarily the brightest neutrino sources. This is explained by the redshift distribution of Fig.~\ref{splitredshiftdistr}. Many of the ULIRGs are found in the tail of the redshift distribution, while LIRGs are found over the whole redshift range. Some ULIRGs therefore have a strong distance-squared suppression, which is not compensated for by their larger IR luminosity. This allows nearby LIRGs to have comparable or larger neutrino flux predictions than ULIRGs.
\\
 
\begin{figure}
    \includegraphics[width=0.45\textwidth]{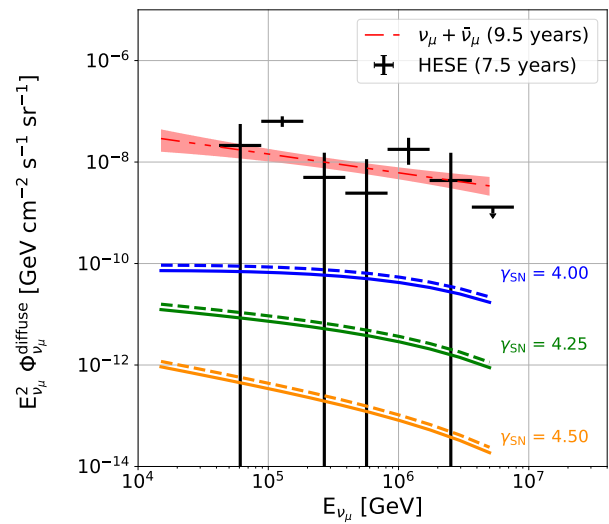}
    \caption{\label{NGOALS}The diffuse starburst-driven muon-neutrino flux expected from 229 disentangled galaxies in GOALS for spectral indices $\gamma_{\mathrm{SN}} = 4.00$, $\gamma_{\mathrm{SN}} = 4.25$, and $\gamma_{\mathrm{SN}} = 4.50$. Note that $\gamma = \gamma_{\mathrm{SN}}-$2. Per spectral index, the diffuse flux is shown with and without correcting the IR luminosity for AGN activity (solid and dashed lines, respectively). For all calculations, $\eta_{\mathrm{tot}} = 0.1 \%$ is used. The black data points are the differential per-flavor IceCube measurements using the high-energy starting event (HESE) sample \cite{Track4}. The red band is the best-fit unbroken power-law spectrum of astrophysical muon neutrinos observed by IceCube in the Northern Hemisphere \cite{Diffusemuon}.}
\end{figure}

Figure \ref{NGOALS} shows the diffuse starburst-driven muon-neutrino flux expected from the GOALS sample for three different spectral indices of the proton injection rate, computed with Eq.$~$(\ref{diffuseGOALS}). For each case, the diffuse flux is shown with and without the use of AGN-corrected IR luminosities indicated by the full lines and dashed lines, respectively. The largest neutrino flux in each case corresponds to the calculations done without correcting for the AGN contribution to the IR luminosity. This increase in flux, observed for all three cases, is driven by the galaxy NGC 1068. It is also concluded from the predictions that none of the parameter combinations violate the diffuse neutrino flux observed by IceCube. Nevertheless, not only local U/LIRGs can contribute to the diffuse neutrino flux, but also the high-redshift counterparts. This is a result of the positive redshift evolution of the comoving IR luminosity density of U/LIRGs. As such, certain combinations of parameter values could still violate the IceCube flux when integrating the U/LIRG contribution over cosmic history.  However, this extrapolation is not trivial for LIRGs as discussed in the next section.

\subsection{\label{Extrapolation} Extrapolation over cosmic history}

Following \cite{Kohta1, Hidden1}, we can obtain an estimate of the diffuse starburst-driven neutrino flux expected from the total LIRG population over cosmic history, based on a representative set of local LIRGs. This representative sample is defined by making a redshift cut on the GOALS sample at $z = 0.0167$ (see Appendix \ref{complete}), which results in 62 disentangled LIRGs and 0 ULIRGs. The integrated cosmic-ray generation rate expected from this sample is assumed to be a fraction $\eta_{\mathrm{tot}}$ of the total IR luminosity generated by the galaxies in that sample. In the context of starburst-driven neutrino production, this fraction is obtained by applying Eq.~(\ref{CRlumformula}) to the local volume under consideration. For typical values  $\eta_{\mathrm{SN}} = 0.10$, $E_{\mathrm{SN}} = 10^{51}$ erg, and $\Lambda_{\mathrm{IR}} = 5.97\times 10^{-46}$ in Eq.~(\ref{CRlumformula}), it is found that $\eta_{\mathrm{tot}}$ $\approx$ 0.1 $\%$. From the integrated cosmic-ray generation rate, the differential rate can be obtained by assuming a spectral index $\gamma$ for the power-law cosmic-ray spectrum at the source (see e.g.~\cite{Hidden1}). The latter in combination with the pp-interaction efficiency $f_{\mathrm{pp}}$ (Sec.~\ref{subsec: calorimete}), fixed to unity for LIRGs, allows to compute the local differential neutrino generation rate. Finally, the diffuse neutrino flux expected from the LIRG population up to redshift $z$ can be obtained by taking into account the redshift evolution factor $\xi_z$ (e.g.~\cite{98Waxman, Pinpoint, ppobscured}). This factor effectively integrates the luminosity function of a source class up to redshift $z$. For an unbroken $E^{-\gamma}$ power-law spectrum of the neutrino emission, the redshift evolution factor becomes independent of energy (e.g.~\cite{ppobscured}) such that

\begin{equation}
        \xi_z = \xi(z,\gamma) = \int^{z}_{0} \frac{\mathrm{d}z'}{\sqrt{\Omega_m(1+z')^3 +\Omega_{\Lambda}}}\mathcal{H}(z)(1+z')^{-\gamma}~,
    \end{equation}

\noindent with $\Omega_m = 0.31$, and $\Omega_{\Lambda}$ = 0.69, and $\mathcal{H}(z)$ the parametrization of the redshift evolution. For the latter we use $\mathcal{H} \propto (1+z)^{m}$ with $m = 4$ for $z \leq 1$ and $m=0$ for $1<z<4$ (e.g.~\cite{Evolution,Pcor,ppobscured}). Integrating up to redshift $z=4$ it follows that $\xi_z = 3.4$ for a spectral index $\gamma = 2$. Taking $\xi_z = 3.4$ into account, we estimate for the diffuse flux from the total LIRG population that $E_{\nu_\mu}^2\Phi^{\mathrm{diffuse}}_{\nu_\mu}=1.95 \times 10^{-8}$ GeV cm$^{-2}$ s$^{-1}$ sr$^{-1}$, which is at the level of the diffuse flux observed with IceCube (see Fig.~\ref{NGOALS}). 

However, pp-interactions at the source produce gamma rays simultaneously with neutrinos. Therefore, any diffuse neutrino flux prediction should be consistent with the nonblazar extragalactic gamma-ray background (EGB) observed with \textit{Fermi} Large Area Telescope (LAT) \cite{Fermidata}. The blazar contribution, which makes up roughly 86$\%$ of the EGB \cite{resolveblazar}, should be subtracted from the total EGB as previous analyses have constrained their contribution to the IceCube flux \cite{FermiBlazar}. Doing so, it has been argued that for a generic cosmic-ray calorimeter scenario the IceCube neutrino flux is in tension with the $\sim$14$\%$ nonblazar EGB between $0.05-1$ TeV \cite{Hidden2}. That is, in such a scenario, the gamma-ray flux expected from the diffuse neutrino flux observed with IceCube overshoots the nonblazar EGB detected by \textit{Fermi}-LAT. 

Since the calorimeter scenario presented in \cite{Hidden2} is generic, it is applicable to any population of hadronuclear neutrino sources that are optically thin to gamma rays in the \textit{Fermi}-LAT energy range. Consequently, as gamma rays are not significantly attenuated in a starburst scenario, our diffuse flux prediction for LIRGs, which is at the level of the IceCube diffuse observations, is likely in tension with the nonblazar EGB. To alleviate this tension, $\gamma > 2$ and/or $\eta_{\rm tot} < 0.1 \%$ could be invoked, as this pushes our predictions below the IceCube flux. It follows that the parameter space of our starburst-driven neutrino-production model is constrained by the nonblazar EGB bound. 

However, the extrapolation presented in this section assumes LIRGs to be standard-candle neutrino emitters. This is an unlikely assumption given the wide range of physical conditions among GOALS U/LIRGs. Considering each source individually could lead to significantly different neutrino flux predictions, as argued in more detail in Sec.~\ref{last}. Furthermore, at $z\sim0$, an IR luminosity cut of $L_{\mathrm{IR}}> 10^{11}L_{\odot}$ favors merger-driven starbursts. At $z \simeq 1-2$, however, when the star-formation rates in galaxies were much higher, $L_{\mathrm{IR}}> 10^{11}L_{\odot}$ targets mostly galaxies that seem to be evolving individually rather than in mergers (e.g.~\cite{localvshighz}). As the physical conditions among LIRGs seem to change with redshift, this could indicate a change in the efficiency of neutrino production with redshift. This should be further investigated as this affects the extrapolation results. Last, we also note that we only considered starburst-driven neutrino production in the extrapolation. However, a fraction of the GOALS U/LIRGs are known to host an (obscured) AGN. Such AGN are promising candidate sources of astrophysical neutrinos as they are typically located in the more central and dusty regions of the galaxy for which significant gamma-ray attenuation is possible. Therefore, obscured AGN could potentially resolve the tension between the diffuse neutrino flux observed by IceCube and the nonblazar EGB observed by \textit{Fermi}-LAT. Particularly interesting U/LIRGs in this context are those known to host Compact Obscured Nuclei (CONs), which are among the most enshrouded regions in the Universe (Sec.~\ref{sec2}). As such, a potential AGN contribution to the neutrino flux should be taken into account in the extrapolation to properly constrain the parameter space of our model. 

\subsection{Per-source vs generic approach\label{last}}

To estimate the diffuse flux predictions in the previous section, all model parameters were fixed in the targeted galaxies except for the IR luminosity and the luminosity distance. However, electromagnetic observations reveal that the fixed model parameters are potentially significantly different among GOALS galaxies. In this section, we highlight the importance of doing per-source investigations to estimate the model parameters and as such properly constrain the neutrino flux of a source.

\begin{figure}
    \includegraphics[width=0.45\textwidth]{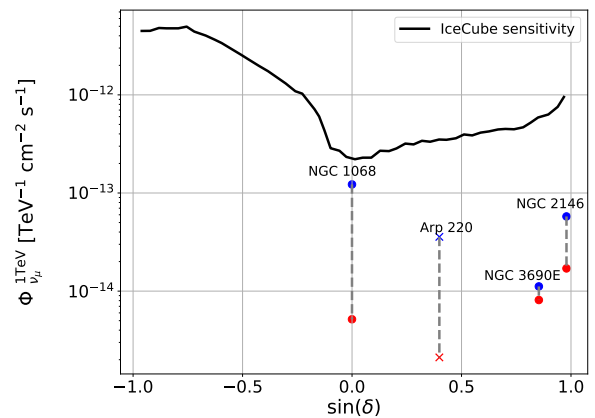}
    \caption{Muon-neutrino flux predictions at 1 TeV as a function of the sine of the declination of four GOALS galaxies identified as gamma-ray sources. For each of the galaxies, the flux prediction is shown for $\gamma_{\mathrm{SN}} = 4$ (blue) and $\gamma_{\mathrm{SN}} = \Gamma + 2$ (red) with $\Gamma$ obtained from gamma-ray observations. These galaxies and their respective $\Gamma$-values are NGC 1068  ($\Gamma = 2.27 \pm 0.09$), NGC 2146 ($\Gamma = 2.27 \pm 0.07$), NGC 3690 ($\Gamma = 2.11 \pm 0.19$), and Arp 220 ($\Gamma = 2.48 \pm 0.14$). For all galaxies $\eta_{\mathrm{tot}} = 0.1 \%$ is used. The symbols have the same meaning as in Fig.~\ref{fig:persource}\label{variablegamma}. }  
\end{figure}

In this section, we consider four GOALS U/LIRGs identified as high-energy gamma-ray sources in \cite{gammaraySFG}. These galaxies and their respective $\Gamma$-values are NGC 1068  ($\Gamma = 2.27 \pm 0.09$), NGC 2146 ($\Gamma = 2.27 \pm 0.07$), NGC 3690 ($\Gamma = 2.11 \pm 0.19$), and Arp 220 ($\Gamma = 2.48 \pm 0.14$). Assuming that these gamma rays are generated by nuclear starburst activity, this hints toward different spectral indices for the neutrino spectra, $\gamma = \gamma_{\mathrm{SN}}-2$. This motivates us to change the spectral index value, while keeping the other model parameters constant, and investigate the effect on the neutrino flux predictions\footnote{Even if these gamma rays are not representative for the neutrino production in the nuclear region, it is still informative to study the effect of changes in $\gamma_{\mathrm{SN}}$.}. Figure \ref{variablegamma} shows the muon-neutrino flux prediction at 1 TeV as a function of the sine of the declination of the investigated gamma-ray sources. Per galaxy, the muon-neutrino flux is shown for $\gamma_{\mathrm{SN}}$ = 4 and for $\gamma_{\mathrm{SN}} = \Gamma +2$. For NGC 1068, $\langle \alpha_{\mathrm{AGN}} \rangle$ is put to zero to compute the neutrino flux. The starburst-driven neutrino flux prediction for NGC 1068 at a given spectral index should therefore be interpreted as an upper limit. Figure \ref{variablegamma} shows that the relative strength of the neutrino flux predictions at $\gamma_{\mathrm{SN}} = 4$ is significantly different from the case where $\gamma_{\mathrm{SN}}$ is variable. Moreover, the flux predictions per source significantly decrease for a spectral index of $\gamma > 2$. As mentioned in the previous section, this could help to resolve the tension between our diffuse neutrino flux prediction and the nonblazar EGB. 

Due to the wide variety in morphologies observed for U/LIRGs (Sec.~\ref{sec:level1}), the average target density encountered by cosmic rays could be significantly different among U/LIRGs. NGC 1068, for example, contains an AGN surrounded by an extended starburst ring while Arp 220 is a merging galaxy in a late stage known to host a much more compact and dense central region. Therefore, the average particle density sampled by cosmic rays could be significantly lower in NGC 1068 than in Arp 220. If the nuclear ISM density encountered by a cosmic ray in NGC 1068 is for example $n=100$ cm$^{-3}$, a factor 10 lower than assumed in Sec.~\ref{PerandDiffuseGOALS}, the expected neutrino flux is a approximately a factor two lower and rapidly decreases for even smaller values of $n$. Furthermore, as the value of $n$ becomes smaller, the neutrino flux predictions become more sensitive to changes in the other model parameters (see Fig.~\ref{H}). It is therefore crucial to understand how cosmic rays propagate and interact within the nuclear ISM. 

The arguments above show the importance of also performing per-source analyses to constrain the neutrino flux from a particular source rather than inferring the latter only from population studies.

\section{Summary}

The extreme IR emission from U/LIRGs traces obscured star formation and AGN activity, which both provide favorable conditions for high-energy neutrino production. In this work, we performed the first investigation of high-energy neutrino emission from LIRGs in GOALS, which is a multiwavelength survey targeting the brightest U/LIRGs in the sky. To do so, we constructed a framework for starburst-driven neutrino production which targets disentangled galaxies in U/LIRG systems. The framework uses the AGN-corrected \textit{Herschel} IR luminosity per galaxy to estimate the cosmic-ray luminosity in that galaxy. Then, by taking into account cosmic-ray propagation in the nuclear region, the neutrino luminosity per U/LIRG can be estimated. The framework requires eight source-specific parameters to compute the expected starburst-driven neutrino flux per galaxy. Each of these parameters were discussed in the context of local U/LIRGs. This study highlighted in a qualitative manner that U/LIRGs are expected to convert high-energy protons into high-energy neutrinos more efficiently than non-U/LIRG starburst galaxies. We then used the framework to:

\begin{itemize}
    \item[--] Estimate the expected neutrino flux generated by the nuclear starburst region in the LIRG NGC 3690. Source-specific electromagnetic data were used to constrain the model parameters whenever possible. From this case study, we concluded that the neutrino flux predictions are most sensitive to changes in the spectral index of the cosmic ray injection rate and the maximum cosmic-ray energy reached in supernova acceleration. These parameters should therefore be the focus of future modeling and experimental efforts. The model predicts that even in the most optimistic cases, the starburst-driven neutrino-flux predictions for NGC 3690 fall one to two orders of magnitude below the current IceCube sensitivity for a source at the declination of NGC 3690. Therefore, these predictions do not violate the IceCube observations. Interestingly, being only an order of magnitude below the current IceCube sensitivity, future observations using extended observatories like IceCube-Gen2 should allow to probe the predicted flux in the more optimistic scenarios.
    
    \item[--] Estimate the diffuse starburst-driven neutrino flux expected from the GOALS sample for different spectral indices. These predictions were found to be orders of magnitude smaller than the diffuse neutrino flux observed by IceCube. Nevertheless, as U/LIRGs have a positive redshift evolution, we also estimated the neutrino flux expected from the total LIRG population across cosmic history. Assuming GOALS LIRGs are standard-candle emitters, we found that cosmic-ray injection spectral indices $\gamma > 2$ and/or infrared conversion efficiencies $\eta_{\mathrm{tot}}<0.1~\%$ are required to avoid tension with the nonblazar extragalactic gamma-ray background observed by \textit{Fermi}-LAT. However, it was also argued that the standard candle assumption for LIRGs is likely unrealistic based on the wide range of nuclear properties observed in LIRGs. Therefore, these population-study results should be interpreted carefully.
    
    \item[--] Estimate the starburst-driven flux expected from NGC 1068. This prediction was compared to the recently reported evidence for a neutrino flux from the direction of NGC 1068 by IceCube. Our flux prediction is significantly smaller than the flux reported by IceCube. Therefore, our model suggests that the neutrino emission from NGC 1068 is likely dominated by an AGN-related process.
\end{itemize}

\section*{Acknowledgements}

This work was supported by the Flemish Foundation for Scientific Research (1149122N, Y. Merckx), the European Unions Horizon 2020 research and innovation program (No 805486, K. D. de Vries), the French National Research Agency (ANR-21-CE31-0025, P. Correa), and the APACHE grant of the French Agence Nationale de la Recherche (ANR-16-CE31-0001, K. Kotera). 

We thank E. Peretti for the valuable feedback and insightful discussions on the cosmic-ray physics. We also thank L. Armus, T. Diaz-Santos, H. Inami, S. Linden, J. Mazzarella, Y. Song, and V. U for their feedback on the U/LIRG aspect of this work.

\appendix

\section{\label{complete} LOCAL LIRG SAMPLE}

GOALS is a subsample of the IRAS RBGS and is therefore a complete flux-limited sample of galaxies with an IRAS 60-$\mu$m flux density of $S_{60 \mu \mathrm{m}, \mathrm{IRAS}} > $ 5.24 Jy. However, GOALS is not a complete volume-limited sample, as suggested by Fig.~\ref{splitredshiftdistr}. To define a representative sample of local LIRGs, we follow the procedure outlined in \cite{Pcor}. First, we estimate the distance up to which the least luminous LIRGs (i.e.$~L_{\mathrm{IR}} = 10^{11} L_{\odot}$) can be detected with the RBGS sensitivity at 60 $\mu\mathrm{m}$, i.e.$~S_{60 \mu \mathrm{m}, \mathrm{IRAS}}  = $ 5.24 Jy. To do so, a fit is performed to the observed correlation between $S_{60 \mu \mathrm{m, IRAS}}$ and the total IR flux, $F_{\mathrm{IR}} = L_{\mathrm{IR}}/4\pi D^2_L$ for all GOALS LIRGs. The fit is of the form 

\begin{equation}
\log_{10}\left(\frac{S_{60\mu \mathrm{m, IRAS}}}{\mathrm{Jy}}\right) = a \log_{10}\left(\frac{F_{\mathrm{IR}}}{\mathrm{W~m^{-2}}}\right) + b~,
\label{loglinearfit}
\end{equation}

 \noindent with best-fit parameters $a = 1.00 \pm 0.01$ and $b = 13.01 \pm 0.08$. Given these parameters, the distance corresponding to $L_{\mathrm{IR}} = 10^{11} L_{\odot}$ and $S_{60 \mu \mathrm{m, IRAS}} = 5.24$ Jy is found by inverting Eq.~(\ref{loglinearfit}). Doing so, we find $D_L \approx 80$ Mpc. In this work, we opt for a conservative value of $D_L = 75$ Mpc which corresponds to $z = 0.0167$. As such, the 62 LIRGs within this redshift define the local sample of LIRGs used to estimate the diffuse neutrino flux from the total LIRG population.

\bibliography{apssamp}

\end{document}